\documentclass[11pt,a4paper]{article}
\usepackage{amsthm,amssymb,amsmath,xy}
\xyoption{all}

  \numberwithin{equation}{section}

\DeclareMathOperator{\End}{\text{End}}
\DeclareMathOperator{\Ker}{\text{Ker}}

\newcommand{\Complex}{\mathbb{C}}

\newcommand{\Natural}{\mathbb{N}}
\newcommand{\Integer}{\mathbb{Z}}

\newcommand{\bartial}{{\bar{\partial}}}
\newcommand{\mat}{\begin{pmatrix}}
\newcommand{\tam}{\end{pmatrix}}
\newcommand{\smat}{\left(\begin{smallmatrix}}
\newcommand{\stam}{\end{smallmatrix}\right)}
\newcommand{\mc}{\mathcal}

\newcommand{\mb}{\mathbb}

\newcommand\Rep{\text{Rep}}
\newcommand\Typ{\text{Typ}}
\newcommand\At{\text{Atyp}}

\newcommand\tr{\text{tr}}

\newcommand\g{\mathfrak{g}}
\newcommand\ag{\mathfrak{\hat{g}}}
\newcommand\h{\mathfrak{h}}

\title{Free fermion resolution of supergroup WZNW models}

\date{}
\author{\\[2mm]Thomas Quella$\,^1$ and Volker Schomerus$\,^2$\\[5mm]\small
$^1$ Korteweg-de Vries Institute for Mathematics,
University of Amsterdam,\\\small
Plantage Muidergracht 24,
1018 TV Amsterdam,
The Netherlands\\[3mm]\small
$^2$ DESY Theory Group, DESY Hamburg,\\\small
Notkestrasse 85, D-22603 Hamburg, Germany
}
\begin{document}

\maketitle
\begin{abstract}
  Extending our earlier work on $PSL(2|2)$, we explain how to reduce
  the solution of WZNW models on general type I supergroups to those
  defined on the bosonic subgroup. The new analysis covers in
  particular the supergroups $GL(M|N)$ along with several close
  relatives such as $PSL(N|N)$, certain Poincar\'e supergroups and the
  series $OSP(2|2N)$. This remarkable progress relies on the use
  of a special Feigin-Fuchs type representation. In preparation
  for the field theory analysis, we shall
  exploit a minisuperspace analogue of a free fermion construction
  to deduce the spectrum of the Laplacian on type I supergroups. The
  latter is shown to be non-diagonalizable. After lifting these
  results to the full WZNW model, we address various issues of
  the field theory, including its modular invariance and the
  computation of correlation functions. In agreement with previous
  findings, supergroup WZNW models allow to study chiral and
  non-chiral aspects of logarithmic conformal field theory within
  a geometric framework. We shall briefly indicate how insights
  from WZNW models carry over to non-geometric examples, such as
  e.g.\ the $\mc{W}(p)$ triplet models.
\end{abstract}

\bigskip

\begin{center}
\begin{tabular}{p{14cm}}
{\bf Keywords:} Conformal Field Theory, Logarithmic Conformal Field Theory,
  Free Field Constructions, Supergroups, Lie Superalgebras,
  Representation Theory
\end{tabular}
\end{center}

\vspace{-17cm} 
\noindent DESY 07-074  \hfill NSF-KITP-07-128 \\
KCL-MTH-07-06 \hfill arXive/0706.0744

\newpage

\tableofcontents

\section{Introduction}

  Two-dimensional non-linear $\sigma$-models on supermanifolds have been a
  topic of considerable interest for the past few decades. Their realm of
  applications is vast, ranging from string theory to statistical
  physics and condensed matter theory. In the Green-Schwarz or pure spinor
  type formulation of superstring theory, for example, supersymmetries
  act geometrically as isometries of an underlying space-time
  (target space) supermanifold. Important examples arise in the
  context of AdS/CFT dualities between supersymmetric gauge theories
  and closed strings. Apart from string theory, supersymmetry has also
  played a major role in the context of quantum disordered systems
  \cite{Parisi:1979ka,Parisi:1982ud,Efetov1983:MR708812,Bernard:1995as}
  and in models with non-local degrees of freedom such as polymers
  \cite{Read:2001pz}. In particular, it seems to be a crucial ingredient
  in the description of the plateaux transitions in the spin
  \cite{Gruzberg:1999dk,Essler:2005ag} and the integer quantum Hall
  effect \cite{Zirnbauer:1999ua,Bhaseen:1999nm,Tsvelik:2007dm}.
\smallskip

  In addition to such concrete applications there exist a number of
  structural reasons to be interested in conformal
  $\sigma$-models with target space (internal) supersymmetry.
  On the one hand, being non-unitary, the relevant conformal
  field theory models exhibit rather unusual features such as
  the occurrence of reducible but indecomposable\footnote{In contrast
  to some appearances in the physics literature we will use the word
  ``indecomposable'' strictly in the mathematical sense. According to
  that definition also irreducible representations are always indecomposable
  since they cannot be written as a direct sum of two other (non-zero)
  representations.} representations and the existence of logarithmic
  singularities on the world-sheet. In this context, many
  conceptual issues remain to be solved, both on the physical and on
  the mathematical side. These include, in particular, the construction
  of consistent local correlation functions \cite{Gaberdiel:1998ps},
  the modular transformation properties of characters
  \cite{Flohr:1995ea,Semikhatov:2003uc}, their relation to fusion rules
  \cite{Fuchs:2003yu,Fuchs:2006nx,Flohr:2007jy}, the treatment of
  conformal boundary conditions \cite{Kawai:2001ur,Gaberdiel:2006pp}
  etc. On the other hand, the special properties of Lie
  supergroups allow for constructions which are not possible for
  ordinary groups. For instance, there exist several families of coset
  conformal field theories that are obtained by gauging a one-sided
  action of some subgroup rather than the usual adjoint
  \cite{Metsaev:1998it,Berkovits:1999zq,Kagan:2005wt,%
  Babichenko:2006uc}. The same class of supergroup $\sigma$-models is
  also known to admit a new kind of marginal deformations that
  are not of current-current type \cite{Bershadsky:1999hk,Gotz:2006qp}.
  Finally, there seems to be a striking correspondence between the
  integrability of these models  and their conformal invariance
  \cite{Bena:2003wd,Young:2005jv,Kagan:2005wt,Babichenko:2006uc}.
\smallskip

  In this note we will focus on the simplest class of two-dimensional
  conformal $\sigma$-models, namely WZNW theories, in order to address
  some of the features mentioned above. The two essential properties
  which facilitate an exact solution are (i) the presence of an extended
  chiral symmetry based on an infinite dimensional current
  superalgebra\footnote{Instead of referring to the names ``Kac-Moody superalgebra''
  or even ``affine Lie superalgebra'' which are frequently used in the
  physics community, we will stick to the notion current superalgebra
  by which we mean a central extension of the loop algebra over a finite
  dimensional Lie superalgebra.} and (ii) the inherent geometric
  interpretation. While (ii) is common to all $\sigma$-models, the
  symmetries of WZNW models are necessary to lift geometric insights
  to the full field theory. Both aspects single out supergroup WZNW
  theories among most of the logarithmic conformal field theories that
  have been considered in the past \cite{Gurarie:1993xq,Gaberdiel:1998ps,%
  Kausch:2000fu} (see also \cite{Gaberdiel:2001tr,Flohr:2001zs} for reviews
  and further references). While investigations of algebraic and mostly
  chiral aspects of supergroup WZNW models reach back more than ten years
  \cite{Rozansky:1992rx,Rozansky:1992td,Maassarani:1996jn,Guruswamy:1999hi,%
  Ludwig:2000em} it was not until recently that the use of geometric methods
  has substantially furthered our understanding of non-chiral issues
  \cite{Schomerus:2005bf,Gotz:2006qp,Saleur:2006tf}. In the last three
  references the full non-chiral spectrum for the $GL(1|1)$, the $PSU(1,1|2)$
  and the $SU(2|1)$ WZNW models has been derived based on methods of harmonic
  analysis. The most important discovery in these articles was the relevance
  of so-called projective covers and the resulting non-diagonalizability of
  the Laplacian which ultimately manifests itself in the logarithmic
  behaviour of correlation functions.
\smallskip

  This paper will put these results on a more general and firm
  conceptual basis by considering rather arbitrary supergroup WZNW models
  based on {\em basic} Lie superalgebras of {\em type~I}. The
  defining properties of these Lie superalgebras are (i) the
  existence of a non-degenerate invariant form (not necessarily
  the Killing form) and (ii) the possibility to split the
  fermionic generators into {\em two} multiplets which
  transform in dual representations of the even subalgebra.
  The first feature is necessary to even spell out
  a Lagrangian for our models. Our second requirement can be
  exploited to introduce a distinguished set of coordinates
  in which the Lagrangian takes a particularly simple form. These
  arise from some Gauss-like decomposition in which a bosonic
  group element is sandwiched between the two sets of fermions.
  The construction resembles the free field construction of
  bosonic models \cite{Feigin1988:MR971497,Gerasimov:1990fi,%
  Bouwknegt:1989jf,Feigin:1990qn,Rasmussen:1998cc}, but its
  fermionic version turns out to be easier to deal with since
  the corresponding Gauss decomposition is globally defined.
  We shall make  no specific choice concerning the coordinates
  on the bosonic subgroup so that the underlying bosonic symmetry is
  manifest throughout our construction.\footnote{See
  \cite{Ito:1992ig,Ito:1992bi} for a related approach.}
  The generators of the underlying current superalgebra of our
  WZNW model are thus constructed from currents of the bosonic
  subalgebra along with a number of free chiral fermionic ghost
  systems which equals the number of fermionic generators.
\smallskip

  As was observed in \cite{Gotz:2006qp} already, at least
  for the example of $PSU(1,1|2)$, the free fermion resolution
  described above provides a natural framework for the discussion
  of representations, spectrum, characters and correlation functions,
  both in the full conformal field theory and in its semi-classical
  subsector. In particular, it is possible to introduce the notion
  of ``Kac modules'' for current superalgebras. These are obtained
  as a tensor product of an irreducible highest weight
  representation of the (renormalized) bosonic subalgebra and a
  Fock space for the free fermions. Exactly as their finite
  dimensional cousins, such Kac modules turn out to be irreducible
  for generic (typical) choices of the highest weight. Hence, their
  characters can be written down immediately and their behaviour
  under modular transformations is straightforward to derive. Though
  our presentation does not allow to elaborate on the details without specifying
  a concrete model, we believe that techniques similar to the ones
  used in \cite{Rozansky:1992td,Gotz:2006qp} also permit to derive
  characters for atypical irreducible representations. The latter
  arise as quotients from reducible Kac modules and their
  characters possess representations through infinite sums over
  characters of Kac modules. Such formulas were shown in
  \cite{Rozansky:1992td,Saleur:2006tf} to provide easy access to
  modular properties of atypical irreducibles.%
\smallskip

  If we restrict our attention to {\em simple} Lie superalgebras for
  a moment our analysis covers three types of infinite series,
  namely $A(m,n)=sl(m|n)$ (for $m\neq n$), $A(n,n)=psl(n|n)$ and
  $C(n+1)=osp(2|2n)$ \cite{Kac:1977em}. But, widening Kac's original
  usage of the qualifiers ``basic'' and ``type~I'', most of our results
  also apply to non-(semi)simple
  Lie superalgebras such as various extended Poincar\'e superalgebras,
  the general linear Lie superalgebras $gl(m|n)$ or supersymmetric
  extensions of Heisenberg algebras.\footnote{WZNW models based on
  Heisenberg algebras may be used to describe strings on maximally
  symmetric plane waves \cite{Nappi:1993ie,D'Appollonio:2003dr}.} We wish to
  stress that our general results below contain a solution of the
  $PSL(n|n)$ WZNW models. What makes these particularly interesting
  is the fact that their volume is an exact modulus, in contrast to
  bosonic non-abelian WZNW models \cite{Bershadsky:1999hk,%
  Babichenko:2006uc}. It is also worth emphasizing that the
  isometries of flat superspace, AdS-spaces and many projective
  superspaces fall into the classes mentioned above. We thus
  expect our work to be relevant for these models as well. A few
  comments in that direction can be found in the conclusions.
\smallskip

  The plan of this paper is as follows. In the next section
  we shall provide a detailed account of Lie superalgebras of type~I
  and the associated representation theory.
  Particular emphasis is put on the structure of projective
  modules, i.e.\ typical Kac modules and projective covers of atypical
  irreducible representations.
  Afterwards we present the supergroup WZNW Lagrangian in
  section~\ref{sc:WZNW} and use a Gauss-like decomposition in
  order to rewrite it in terms of a bosonic WZNW model, two sets
  of free fermions and an interaction term which couples bosons and
  fermions. This free fermion resolution is shown to have an
  algebraic analogue on the level of the current superalgebra which
  constitutes the symmetry of the supergroup WZNW model.
  The analysis of the zero-mode spectrum in the large
  volume sector is performed in section~\ref{sc:SCA} using methods
  of harmonic analysis. Most importantly, we shall determine the
  representation content for the combined left right regular action
  on the algebra of functions over the supergroup. To achieve
  our goal, we use a reciprocity between atypical irreducible
  representations and their projective covers. On the way we also
  prove the occurrence of a sector on which the Laplacian is not
  diagonalizable. After these preparations we extend the free
  fermion resolution to the full WZNW model in section~\ref{sc:Quantum}.
  We introduce an analogue of Kac modules suitable for infinite
  dimensional Lie superalgebras and sketch the calculation of
  correlation functions. The latter are necessarily logarithmic
  due to the non-diagonalizability of the dilatation operators
  $L_0$ and $\bar{L}_0$. At the end of section~\ref{sc:Quantum},
  we propose a universal
  partition function resembling a charge conjugate invariant and
  gather some thoughts about the possibility of having non-trivial
  modular invariant partition functions. In the concluding
  section~\ref{sc:triplet} we argue that the solution of the
  logarithmic triplet model \cite{Gaberdiel:1998ps} formally fits
  into the framework outlined before. This observation is used to
  speculate about the structure of general logarithmic conformal
  field theories.%
\smallskip

  Most of the statements which appear in the main text can
  be turned into mathematically rigorous propositions. This applies
  in particular to all algebraic manipulations. In our discussion of
  spectra, however, we focus on models based on finite dimensional
  representations. The most interesting supergroups, on the other hand,
  are based on {\em non-compact} and occasionally on {\em non-reducive}
  groups. While we believe that our discussion may be
  extended to such cases, a fully comprehensive presentation
  would have required to carefully distinguish between different real
  forms. In the present note, we rather preferred to put the emphasis
  on the algebraic structures that -- in our opinion -- are equally
  relevant for {\em all} type~I supergroup WZNW models.

\section{Some background on Lie superalgebras of type~I}

  The main actress of this paper, the Lie supergroup $G$, is best introduced
  in terms of its underlying Lie superalgebra $\g$. We
  will assume the latter to be finite dimensional, basic and of type~I. The
  attribute ``basic'' guarantees the existence
  of a non-degenerate invariant metric and is needed in order to exclude
  certain pathological cases which would even rule out the existence of a
  WZNW Lagrangian. The predicate ``type~I'', on the
  other hand, implies the existence of two multiplets of fermionic
  generators and will simplify the interpretation of the chiral
  splitting in the conformal field theory we are considering.
\smallskip

  In the remainder of this section we shall first present the
  commutation relations of a general (possibly non-simple) basic
  Lie superalgebra of type~I. Afterwards we summarize their representation
  theory following the beautiful exposition of Zou \cite{Zou1996:MR1378540}
  (see also \cite{Brundan2004:MR2100468}). The reader who is not interested
  in the mathematical details might wish to skip over parts of this section
  in the first reading.

\subsection{\label{sc:CR}Commutation relations}

  A Lie superalgebra $\g=\g_0\oplus\g_1$ is a graded generalization
  of an ordinary Lie algebra \cite{Kac:1977em}. There are even (or
  {\em bosonic}) generators $K^i$ which form an ordinary Lie algebra $\g_0$,
  i.e.\ they obey the commutation relations
\begin{equation}
  \label{eq:CommBB}
  [K^i,K^j]\ =\ i{f^{ij}}_l\,K^l\ \ ,
\end{equation}
  with structure constants that are antisymmetric in the upper indices
  and that satisfy the Jacobi identity.
  In addition, type~I Lie superalgebras possess two
  sets of odd (or {\em fermionic}) generators $S_1^a$ and $S_{2a}$, $a=1,\ldots,r$
  (generating $\g_1$) which transform in an $r$-dimensional
  representation $R$ of $\g_0$ and its dual $R^\ast$, respectively
  \cite{Kac1978:MR519631}. Rephrased
  in terms of commutation relations, this statement may be expressed as
\begin{align}
  \label{eq:CommBF}
  [K^i,S_1^a]&\ =\ -{(R^i)^a}_b\,S_1^b&
  [K^i,S_{2a}]&\ =\ {S_{2b}\,(R^i)^b}_a\ \ .
\end{align}
  The symbol $R^i$ is an abbreviation for the representation matrix
  $R(K^i)$. In a type~I superalgebra, the anti-commutators $[S_1^a,S_1^b]$
  and $[S_{2a},S_{2b}]$ vanish identically \cite{Kac1978:MR519631}. On the
  other hand, generators $S_1^a$ do not anti-commute with $S_{2b}$. Before
  we are able to spell out their commutation relations, we need to
  introduce the supersymmetric bilinear form
\begin{align}
  \label{eq:Metric}
  \langle K^i,K^j\rangle&\ =\ \kappa^{ij}&
  \langle S_1^a,S_{2b}\rangle&\ =\ \delta_b^a\ \ .
\end{align}
  We assume $\kappa^{ij}$ (not necessarily the Killing form) to be
  invariant with respect to $\g_0$ and, moreover, to be non-degenerate
  such that its inverse $\kappa_{ij}$ exists. The latter is a crucial
  ingredient in the definition
\begin{equation}
  \label{eq:CommFF}
  [S_1^a,S_{2b}]\ =\ -{(R^i)^a}_b\,\kappa_{ij}\,K^j\ \ .
\end{equation}
  The structure constants which appear in this relation are uniquely
  determined by the requirement that the metric \eqref{eq:Metric} is
  invariant, i.e.\ $\langle[X,Y],Z\rangle=\langle X,[Y,Z]\rangle$.
  The supersymmetry and non-degeneracy of the metric on the full Lie
  superalgebra $\g$ follow immediately from the definition.%
\smallskip

  The commutation relations above preserve the fermion number $\#(S_1^a)-\#(S_{2a})$. 
  Hence $\g$ and also its universal enveloping superalgebra $\mc{U}(\g)$ have a
  natural $\Integer$-grading (localized in three degrees) which is
  consistent with the intrinsic $\Integer_2$-grading \cite{Kac1978:MR519631}.
  It is this property which distinguishes type~I Lie superalgebras among all
  Lie superalgebras. Let us also emphasize that the Cartan
  subalgebra of $\g$ will always be identified with that of
  $\g_0$ in what follows. This will be important below when we
  introduce highest weight representations.
\smallskip

  Before we end this subsection on the definition of type~I
  superalgebras, let us reflect a bit on how restrictive their
  structure is. In fact, in
  building a Lie superalgebra one cannot just come up with any bosonic
  subalgebra $\g_0$ and hope to extend it by adding fermions transforming
  in some representation $R$ of $\g_0$. There is an additional
  constraint, namely the graded Jacobi identity. While the latter
  is by assumption identically satisfied for $\g_0$ and the
  mixed bosonic/fermionic commutators do not impose any new
  conditions, there is a non-trivial restriction arising from
  the commutator $[S_1^a,[S_1^b,S_{2c}]]$ and its cyclic permutations.
  This leads to the requirement
\begin{equation}
  \label{eq:Jacobi}
  {(R^i)^b}_c\,\kappa_{ij}\,{(R^j)^a}_d+{(R^i)^a}_c\,
   \kappa_{ij}\,{(R^j)^b}_d\ =\ 0\ \ .
\end{equation}
  An equivalent formulation is to demand that the quadratic Casimir
  vanishes on the symmetric part of the tensor product $R\otimes R$.
  Alternatively, the constraint on the choice of $R$ may be rephrased
  by requiring that the tensor
\begin{equation} \label{Rch2}
  {A^{ab}}_{cd}\ =\ {(R^i)^a}_c\,\kappa_{ij}\,{(R^j)^b}_d
\end{equation}
  is antisymmetric in the upper two as well as the lower two
  indices. Although the property \eqref{eq:Jacobi} (or \eqref{Rch2})
  looks rather innocent it will be a crucial ingredient in many of the
  equalities we shall encounter.

\subsection{\label{sc:RepTh}Representation theory}

  In the analysis of supergroup WZNW models there are a variety
  of representations of the underlying Lie superalgebra which
  play a role. The aim of this section is to provide a brief summary
  of the relevant modules of finite dimensional type~I Lie
  superalgebras following Zou's exposition \cite{Zou1996:MR1378540}.
  For definiteness, all the definitions and statements that follow
  below will be formulated for finite dimensional representations.
  It is understood, though, that our definitions can be extended to
  infinite dimensional representations (discrete and continuous)
  as well. Whether this is also true for their properties, however,
  remains to be investigated.

\subsubsection{\label{ssc:KacMod}Kac modules and their duals}

  Let us denote by $\Rep(\g_0)$ the set of isomorphism classes of
  irreducible representations of the bosonic subalgebra $\g_0$.
  The basic building blocks in the representation theory of Lie superalgebras
  of type~I are {\em Kac modules} $\mc{K}_\mu$, $\mu\in\Rep(\g_0)$
  \cite{Kac:1977em,Kac1978:MR519631}.
  They are induced from irreducible representations $V_\mu$ of the
  bosonic subalgebra $\g_0$. More precisely, the representation is
  extended by letting one multiplet of fermionic generators $S_{2a}$
  act trivially on the vectors $v\in V_\mu$. The remaining states in
  the Kac module are then created by acting with generators from the
  second multiplet of fermions, $S_1^a$. From our verbal description
  we immediately infer the decomposition of Kac modules with respect
  to the bosonic subalgebra,\footnote{In the following
  we shall refer to the right hand side of this equation and
  other restrictions of $\g$-modules to the bosonic subalgebra
  $\g_0$ as the ``bosonic content''. Hence, this phrase is not
  related to the $\Integer_2$-grading of the representation
  space.}
\begin{equation}
  \mc{K}_\mu\bigr|_{\g_0}
  \ =\ V_\mu\otimes\mc{F}
  \ =\ {\bigoplus}_\nu\,\bigl[\mc{K}_\mu:V_\nu\bigr]_0\,V_\nu\ \ .
\end{equation}
  Here and in what follows we assume all $\g_0$-modules to be fully
  reducible and denote the resulting multiplicities in terms of
  the square bracket $[M_0:N_0]_0$ where $M_0$ is an arbitrary (fully
  reducible) $\g_0$-module and $N_0$ an irreducible $\g_0$-module.
  The $\g_0$-module $\mc{F}=\bigwedge(S_1^a)$ appearing in the
  previous equation is the exterior (or Grassman) algebra generated
  by the fermions $S_1^a$. Its structure as a $\g_0$-module is
  determined by projecting tensor powers of the module $R^\ast$
  onto their fully anti-symmetric submodules,
\begin{equation}
  \label{eq:FermRep}
  \mc{F}\ =\ V_0^\ast\oplus R^\ast\oplus\bigl[R^\ast\otimes R^\ast\bigr]_{\text{antisym}}
             \oplus\bigl[R^\ast\otimes R^\ast\otimes R^\ast\bigr]_{\text{antisym}}
             \oplus\cdots\oplus\bigl[(R^\ast)^{\otimes r}\bigr]_{\text{antisym}}\ \ .
\end{equation}
  The $n$-fold tensor product here corresponds to a state
  involving $n$ fermionic generators $S_1^{a_i}$, $i=1,\ldots,n$.
  The case of no fermionic generators leads to
  the one-dimensional trivial representation $V_0=V_0^\ast$.
  It is obvious that the series will truncate after the $r$-th
  tensor product since the fermionic generators $S_1^a$ anti-commute
  among themselves. Consequently, the dimension of Kac modules
  is always given by $\dim(\mc{K}_\mu)=2^r\dim(V_\mu)$.
\smallskip

  In close analogy to the previous definition we may also
  introduce {\em dual Kac modules} $\mc{K}_\mu^\ast$ by starting
  with the dual bosonic representation $V_\mu^\ast=V_{\mu^+}$.
  Deviating from the above construction we now let the first
  set of fermionic generators $S_1^a$ act trivially on the
  corresponding vectors and use $S_{2a}$ to create new states.
  Since the two sets of fermionic generators transform in dual
  representations the bosonic content is then obviously given by
\begin{equation}
  \mc{K}_\mu^\ast\bigr|_{\g_0}
  \ =\ V_\mu^\ast\otimes\mc{F}^\ast
  \ =\ \bigl(V_\mu\otimes\mc{F}\bigr)^\ast
  \ =\ {\bigoplus}_\nu\,\bigl[\mc{K}_\mu^\ast:V_\nu\bigr]_0\,V_\nu\ \ .
\end{equation}
  The dimensions of the modules $\mc{K}_\mu$ and
  $\mc{K}_\mu^\ast$ coincide and it may easily be seen that
  the representations are indeed dual to each other.
\smallskip

  Let us conclude the discussion of Kac modules with a short comment about
  the last term in the fermionic representation $\mc{F}$, eq.\ \eqref{eq:FermRep}.
  Innocent as it seems, it is important to stress that the highest
  component $[R^{\otimes r}]_{\text{antisym}}$ need not be the
  trivial $\g_0$-module $V_0$ again, even though it certainly is
  one-dimensional. The action of the bosonic subalgebra on this
  space can be calculated explicitly,
\begin{equation}
  K^i\cdot\bigl(S_1^1\cdots S_1^r\bigr)\ =\ -\tr(R^i)\,S_1^1\cdots S_1^r\ \ .
\end{equation}
  In case $\g_0$ is semisimple, it admits a unique one-dimensional
  representation, namely the trivial $\g_0$-module $V_0$. Hence, we conclude
  that $\tr(R^i)=0$ for Lie superalgebras with a semisimple bosonic subalgebra.

\subsubsection{Simple modules and their blocks}

  Kac modules provide an important intermediate step to
  constructing {\em irreducible representations}. Finding their exact
  relation with irreducibles, however, requires good control over
  the structure of Kac modules. For generic labels $\mu$, the
  (dual) Kac modules turn out to be irreducible. Thereby, they
  give rise to what is known as {\em typical} irreducible
  representations $\mc{L}_\mu=\mc{K}_\mu$. But there exist special
  values of $\mu$ for which the associated Kac module contains a
  proper invariant subspace. The so-called {\em atypical} irreducible
  representations $\mc{L}_\mu$ are obtained from such $\mc{K}_\mu$ by
  factoring out the unique maximal invariant submodule
  \cite{Kac1978:MR519631}. In contrast to
  the typical case, it is not straightforward to give a general formula
  for the dimension or the bosonic content of atypical irreducible
  representations, see however \cite{Zou1996:MR1378540,Serganova1998:MR1648107}.
  As will be explained below the representations $\mc{L}_0$ as well as $\mc{L}_R$
  and $\mc{L}_R^\ast=\mc{L}_{R^\ast}$ are always atypical.
\smallskip

  We shall assume that all irreducible representations of our type~I
  superalgebra $\g$ emerge as (possibly trivial) quotients of Kac modules
  (cf.~\cite{Kac1978:MR519631}). In other words, the set $\Rep(\g)$
  of isomorphism classes (or labels) of irreducible $\g$-modules
  agrees with the one of its bosonic subalgebra, i.e.\ $\Rep(\g)=
  \Rep(\g_0)$. According to our previous remarks, it splits into
  two disjoint sets, $\Rep(\g)=\Typ(\g)\cup\At(\g)$, containing
  typical and atypical labels, respectively.
\smallskip

  Simple modules of a Lie superalgebra can be grouped into so-called
  {\em blocks}. By definition, blocks are the parts of the finest
  partition of $\Rep(\g)$ such that
  two simple modules belong to the same part as soon as they have a
  non-split extension (see, e.g., \cite{Germoni1998:MR1659915}).
  An intuitive way of understanding this definition is to view
  the simple modules as vertices in a graph. There exists an edge
  between two vertices if and only if the corresponding simple
  modules admit a non-split extension. In this picture,
  the blocks correspond to connected components of the full graph.
  The property ``being connected'' defines
  an equivalence relation $\sim$ on $\Rep(\g)$. We will use
  the notation $\Gamma(\g)=\Rep(\g)/\sim$ for the set of all blocks
  and $[\sigma]\in\Gamma(\g)$ for individual blocks. Notice that
  each typical module forms a block by itself.\footnote{This statement
  only holds in this form if we restrict ourselves to finite dimensional
  representations.} Atypical irreducible representations, on the
  other hand, form constituents of larger blocks. This implies
  the decomposition
  $\Gamma(\g)=\Gamma_{\text{typ}}(\g)\cup\Gamma_{\text{atyp}}(\g)$
  where $\Gamma_{\text{typ}}(\g) = \Typ(\g)$.
\smallskip

  It is easy to argue that each Lie superalgebra of type~I
  possesses a (probably infinite) block $[0]$ containing
  the trivial representation. Atypicality of the one-dimensional
  trivial representation already follows on dimensional
  grounds since the dimension of Kac modules is always a multiple of
  $2^r$. Let us continue to show that the representations $\mc{L}_R$
  and $\mc{L}_R^\ast\cong\mc{L}_{R^\ast}$ which are based on
  the $\g_0$-modules $R$ and $R^\ast$ lie in the same block
  $[0]$. It is straightforward to see that $\mc{L}_0$
  is obtained as a quotient from the Kac module $\mc{K}_0$
  where the subscript $0$ refers to the trivial $\g_0$-module.
  In order to prove the atypicality of $\mc{L}_R$ we consider
  the states in $\mc{K}_0$ which are obtained from the ground state
  by applying precisely one fermionic generator. These states
  transform in the representation $R$ of $\g_0$. Since the
  Kac module $\mc{K}_0$ is atypical and its irreducible quotient is
  of dimension one, this representation has to
  decouple, i.e.\ the fermionic generators $S_{2a}$ have to annihilate
  these states. We observe that the representation $R$ can be
  part of at least two different supermultiplets: it may be used to
  define a Kac module $\mc{K}_R$ and it generates a submodule $\mc{Q}_R$
  of $\mc{K}_0$. In both cases, the highest weight conditions are exactly
  identical. But obviously the dimensions of $\mc{Q}_R$ and $\mc{K}_R$ do
  not coincide since $\dim\mc{Q}_R<\dim\mc{K}_0<\dim\mc{K}_R$.
  Hence, $\mc{Q}_R$ has to be a non-zero quotient of $\mc{K}_R$,
  proving the atypicality of the latter. The same reasoning could
  be repeated with at least one of the $\g_0$-modules which appear in the
  (dual) Kac modules $\mc{K}_R$ and $\mc{K}_{R^\ast}$ and so on. Thereby we
  construct a presumably infinite chain of atypical representations
  $\mc{L}_\mu$ in the block $[0]$. The labels that are included in
  this block all appear in the decomposition of the tensor products
  $R^{\otimes m}\otimes(R^\ast)^{\otimes n}$ for arbitrary powers
  $m$ and $n$ (the converse is not true, of course).

\subsubsection{\label{sc:ProjMod}Projective modules}
  Lie superalgebras possess a whole zoo of representations
  which cannot be decomposed into a direct sum of irreducibles. We shall
  see some important examples momentarily. Let us recall before that any
  $\g$-module $M$ possesses a {\em composition series}. The latter is
  determined by a special kind of filtration, in the present case an
  ascending set of submodules $M_i$, $i=0,\ldots,n$ where $M_0=0$
  and $M_n=M$, such that the quotients $M_i/M_{i-1}$ are simple
  modules. We will denote by $[M:\mc{L}_\mu]$ the number of
  irreducible $\g$-modules $\mc{L}_\mu$ in this composition series
  of $M$.
\smallskip

  The most interesting class of indecomposables consists of the
  so-called {\em projective covers} $\mc{P}_\mu$ of irreducibles
  $\mc{L}_\mu$. The module $\mc{P}_\mu$ is defined to be the unique
  indecomposable projective module that contains the irreducible
  representation $\mc{L}_\mu$ as its head.\footnote{The attribute
  ``projective'' is used here in the sense of category theory and
  should not be confused with the notion of projective representations
  that is used when algebraic relations are only respected up to some
  multipliers (cocycles).} By definition, the head of a representation
  is the quotient by its maximal proper submodule.
  For typical labels one has the equivalences
  $\mc{L}_\mu\cong\mc{K}_\mu\cong\mc{P}_\mu$. For atypical labels,
  however, irreducible modules, Kac modules and projective covers
  are all inequivalent. In particular, they possess different
  dimensions.
\smallskip

  All projective modules $\mc{P}$ of a type~I superalgebra are known
  to possess a {\em Kac composition series} \cite{Zou1996:MR1378540},
  i.e.\ a filtration in terms of submodules whose quotients are Kac
  modules.\footnote{It should be stressed that this property is
  not true for type~II Lie superalgebras. A counter-example is provided
  by $D(2,1;\alpha)$ whose representation category is discussed in
  \cite{Germoni2000:MR1840448}.}
  We denote by $(\mc{P}:\mc{K}_\lambda)$ the number of Kac
  modules $\mc{K}_\lambda$ in the Kac composition series of
  $\mc{P}$. In order to describe the precise
  structure of indecomposable projective modules we will rely on the
  following reciprocity theorem \cite[Theorem 2.7]{Zou1996:MR1378540}
  (see also \cite{Brundan2004:MR2100468})
\begin{equation}
  \label{eq:BGG}
  \bigl(\mc{P}_\mu:\mc{K}_\lambda\bigr)\ =\ \bigl[\mc{K}_\lambda:\mc{L}_\mu\bigr]\ \ .
\end{equation}
  This important equation relates the multiplicities of Kac modules
  in the Kac composition series of a projective cover to the multiplicity
  of irreducible representations arising in the composition series of Kac
  modules. Hence, the structure of projective covers is completely
  determined by that of Kac modules. The statement is trivial for
  typical labels but it contains valuable information in the
  atypical case. Note that a small technical assumption underlying Zou's
  proof of eq.~\eqref{eq:BGG} seems to be overcome if one uses the
  approach of \cite{Brundan2004:MR2100468}.
\smallskip

  There is one simple construction that is guaranteed to furnish
  projective modules and it is exactly this construction
  through which the latter will enter in our harmonic analysis later on.
  The idea is to induce representations from irreducible representations
  $V_\mu$ of $\g_0$ by letting {\em both} sets of fermionic generators
  $S_1^a$ and $S_{2a}$ act non-trivially, i.e.\
\begin{equation}
  \mc{B}_\mu\ =\ \text{Ind}_{\g_0}^{\g}(V_\mu)\ \ .
\end{equation}
  These modules are projective and reducible \cite{Zou1996:MR1378540}.
  Indeed, under reasonable assumptions on $\g_0$ all finite dimensional
  $\g_0$-modules are projective, and this property is preserved by
  induction. For later use, let us write down the decomposition of the
  representations $\mc{B}_\mu$ into their indecomposable building
  blocks. We start with the observation that their bosonic content
  is given by
\begin{equation}
  \mc{B}_\mu\bigr|_{\g_0}\ =\ V_\mu\otimes\mc{F}\otimes\mc{F}^\ast\ \ .
\end{equation}
  Using a suitable rearrangement of these factors it is obvious
  that the multiplicities of Kac modules in the Kac composition series
  of $\mc{B}_\mu$ are given by
  $(\mc{B}_\mu:\mc{K}_\nu)=[\mc{K}_{\mu^+}^\ast:V_\nu]_0$.
  For the actual decomposition into indecomposables
  we use our knowledge that $\mc{B}_\mu$ is projective. This implies that
  it may be written as a direct sum of (typical) irreducible Kac modules
  and (atypical) projective covers. While nothing remains to be
  done for typical representations, the correct description of the atypical
  sector requires combining the corresponding (non-projective) Kac modules
  into projective covers. In order to achieve this goal we note the
  equality $[\mc{K}_{\mu^+}^\ast:V_\nu]_0=[\mc{K}_\nu^\ast:V_{\mu^+}]_0$
  which holds because both sides correspond to the number of
  $\g_0$-invariants in the tensor product $V_\mu\otimes V_\nu^\ast\otimes
  \mc{F}^\ast$. Now we can use the following simple consequence of
  the duality relation \eqref{eq:BGG},
\begin{equation}
  [\mc{K}_{\mu^+}^\ast:V_\nu]_0
  \ =\ \bigl[\mc{K}_\nu^\ast:V_{\mu^+}\bigr]_0
  \ =\ {\sum}_\sigma\,\bigl[\mc{K}_\nu^\ast:\mc{L}_\sigma^\ast\bigr]\,
  \bigl[\mc{L}_\sigma^\ast:V_{\mu^+}\bigr]_0
  \ =\ {\sum}_\sigma\,\bigl(\mc{P}_{\sigma}:\mc{K}_\nu\bigr)\,
  \bigl[\mc{L}_\sigma^\ast:V_{\mu^+}\bigr]_0\ \ ,
\end{equation}
  to arrive at the final result
\begin{equation}
  \label{eq:BStructure}
  \mc{B}_\mu\ =\ \bigoplus_{\nu\in\Typ(\g)}\bigl[\mc{K}_{\mu^+}^\ast:
   V_\nu\bigr]_0\,\mc{K}_\nu
                 \oplus\bigoplus_{\sigma\in\At(\g)}\bigl[\mc{L}_\sigma^\ast:
  V_{\mu^+}\bigr]_0\,\mc{P}_\sigma\ \ .
\end{equation}
  This formula will be one of the main ingredients in the
  harmonic analysis to be performed below in section \ref{sc:HarmAn}.
  It is interesting to note that {\em every} indecomposable
  projective module arises as a subspaces of some
  $\mc{B}_\mu$ \cite{Zou1996:MR1378540}. This means that the category
  of representations considered here ``contains enough projectives''.
\smallskip

  Let us elaborate a bit more
  on the distinguished role that projective modules -- direct sums of typical
  irreducibles and projective covers of atypical irreducibles -- play
  for the representation theory of Lie superalgebras. In fact, in many ways
  they take over the role of irreducible representations in the
  theory of ordinary Lie algebras. Most importantly, it can be shown
  that the tensor product of any module with a projective one is
  projective again. In other words, projective modules form an ideal
  in the representation ring. Moreover, the Clebsch-Gordon decomposition
  for tensor products of projective modules can be determined through
  a variant of the Racah-Speiser algorithm. Consider for instance two
  projective $\g$-modules $A_1$ and $A_2$. Being projective, they have
  a Kac composition series and hence their bosonic content is
  given by
\begin{equation}
  A_i\bigr|_{\g_0}\ =\ {\sum}_{\mu}m_{i\mu}V_\mu\otimes\mc{F}\ \ .
\end{equation}
  For the bosonic content of the tensor product $A_1 \otimes A_2$
  this implies
\begin{equation}
  \bigl(A_1\otimes A_2\bigr)\bigr|_{\g_0}
  \ =\ {\sum}_{\mu}m_{1\mu}m_{2\nu}\Bigl[V_\mu\otimes V_\nu\otimes
  \mc{F}\Bigr]\otimes\mc{F}\ \ .
\end{equation}
  The last $\mc{F}$ should be interpreted as the fermionic factor
  that is guaranteed to be present in every projective module, due
  to the fact that they possess a Kac composition series. All we
  need to do is to decompose the factor $V_\mu\otimes
  V_\nu\otimes\mc{F}$ into irreducibles of $\g_0$. This provides
  us with a list of all Kac modules in $A_1 \otimes A_2$ along
  with their multiplicities. Typical Kac modules correspond
  to irreducible representations appearing in the tensor product
  while atypical Kac modules must be re-combined into projective
  covers. This final step is performed based on formula~\eqref{eq:BGG}
  and it leads to an unambiguous result. Our discussion shows how the
  Clebsch-Gordon decomposition of the
  tensor product $A_1 \otimes A_2$ may be played back to the
  bosonic subalgebra. The decomposition of $V_\mu\otimes
  V_\nu\otimes\mc{F}$ can be tackled with the usual algorithmic
  tools from the representation theory of Lie algebras.

\subsubsection{The quadratic Casimir element}
  One of the most important objects in representation theory are the
  {\em Casimir elements}, i.e.\ elements of the center of the
  universal enveloping algebra $\mc{U}(\g)$. For our concrete choice
  of generators and invariant form we have a natural quadratic Casimir
\begin{equation}
  \label{eq:Casimir}
  C\ =\ K^i\kappa_{ij}K^j-S_1^aS_{2a}+S_{2a}S_1^a\ \ .
\end{equation}
  It may easily be checked that this operator acts as
  a scalar on Kac modules $\mc{K}_\mu$. For a vector $v\in V_\mu$
  in the defining irreducible bosonic multiplet one finds
\begin{equation}
  \label{eq:Eigenval}
  Cv\ =\ \bigl(C_B-\tr(R^i)\,\kappa_{ij}\,K^j\bigr)v\ \ ,
\end{equation}
  where $C_B=K^i\kappa_{ij}K^j$ is the quadratic Casimir of $\g_0$
  associated to its non-degenerate metric. Since
  the second term inside the bracket commutes with $\g_0$ as well, the
  irreducibility of $V_\mu$ implies that $C$ acts as a scalar on
  the whole multiplet $V_\mu$. Using the commutativity of $C$
  with $\g$, this action may be extended to the complete Kac module
  $\mc{K}_\mu$. We will denote the corresponding eigenvalue
  of the Casimir by $C_\mu=C(\mc{K}_\mu)$. Because irreducible
  $\g$-modules are defined as a quotient of Kac modules this
  immediately implies $C(\mc{L}_\mu)=C(\mc{K}_\mu)$.
\smallskip

  The observation that several representations may have the same
  Casimir eigenvalues can be seen to generalize. In fact, it just takes
  a moment of thought to convince oneself that one has $C_\mu=C_\nu$
  (and the same for other Casimirs) whenever the simple modules
  belong to the same block, $\mu\sim\nu$. It seems plausible that
  also the converse holds, i.e.\ that the set of Casimir operators
  may be used to separate different blocks. If this assertion was
  true, then choosing $\mu$ and $\nu$ from different blocks, one
  would be able to find a Casimir (not necessarily quadratic)
  whose eigenvalues on $\mc{L}_\mu$ and $\mc{L}_\nu$ disagree.
\smallskip

  The previous comment that Casimir eigenvalues are constant
  on blocks has interesting implications for indecomposables.
  By definition, the composition series of an indecomposable
  contains irreducibles belonging to one and the same block. Therefore,
  within any indecomposable, no matter how complicated it is, all
  generalized eigenvalues of the Casimir elements are the same.
  The additional qualifier ``generalized'' is necessary because
  a Casimir element need not be diagonalizable when evaluated
  in an indecomposable representation. This phenomenon is
  particularly common for the projective covers of atypicals.
  We shall see later that -- at least for a type~I Lie superalgebra
  -- the quadratic Casimir \eqref{eq:Casimir} cannot be diagonalized
  in any of the projective covers $\mc{P}_\mu$.\footnote{Diagonalizability
  might be true for other Casimir operators though. For $gl(1|1)$, for
  example, the Casimir element $E^2$ is diagonalizable in all weight
  modules. Note however that $E^2$ is not related to a {\em non-degenerate}
  invariant form as in eq.\ \eqref{eq:Casimir}.}
  Furthermore, there exists at least one series of projective covers,
  the ones associated to the block $[0]\in\Gamma(\g)$ of the trivial
  representation, for which the generalized eigenvalues, i.e.\ the
  diagonal entries in the Jordan block, vanish
  identically.\footnote{Certain type~II superalgebras
  such as e.g.\ $D(2,1;\alpha)$ are known to also possess projective
  covers with non-vanishing generalized eigenvalues
  \cite{Germoni2000:MR1840448}.}

\section{\label{sc:WZNW}The supergroup WZNW model and its symmetries}

  In this section we will introduce the WZNW model using its Lagrangian
  formulation. We will employ a Gauss-like decomposition in order to
  rewrite the Lagrangian in terms of a bosonic WZNW model, a free fermion
  theory and an interaction term. We then describe the infinite dimensional
  current superalgebra of the model and explain how the latter may
  be reconstructed from the free fermion resolution introduced before.
  Let us stress that, contrary to the well known free field approaches
  to bosonic WZNW models
  \cite{Feigin1988:MR971497,Gerasimov:1990fi,Bouwknegt:1989jf,Feigin:1990qn,%
  Rasmussen:1998cc}, our approach keeps the full bosonic symmetry manifest at
  all times. It reduces the problem of solving the supergroup WZNW model
  to a solution of the underlying bosonic model.

\subsection{\label{sc:Lagrange}The Lagrangian description}

  Given the Lie superalgebra $\g$ as defined in
  \eqref{eq:CommBB}-\eqref{eq:CommFF}, we can combine its generators
  with elements of a Grassmann algebra in order to obtain a Lie algebra
  which can be exponentiated. In physicist's manner we shall
  define the supergroup $G$ to be given by elements
\begin{equation}
  \label{eq:Para}
  g\ =\ e^{\theta}\,g_B\,e^{\bar{\theta}}
\end{equation}
  with $\theta=\theta^aS_{2a}$ and $\bar{\theta}=\bar{\theta}_bS_1^b$
  (this parametrization has been termed ``chiral superspace'' in
  \cite{Gates:1983nr}). The coefficients $\theta^a$ and
  $\bar{\theta}_b$ are independent Grassmann
  variables while $g_B$ denotes an element of the bosonic
  subgroup $G_B\subset G$ obtained by exponentiating the
  Lie algebra generators $K^i$. The attentive reader may have
  noticed that the product of two such supergroup elements
  \eqref{eq:Para} will not again give a supergroup
  element of the same form. We shall close an eye on such
  issues. For us, passing through the supergroup is merely an
  auxiliary step that serves the purpose of constructing a
  WZNW-like conformal field theory with Lie superalgebra
  symmetry. Since Lie superalgebras do not suffer from
  problems with Grassmann variables, the resulting conformal
  field theory will be well-defined.
\smallskip

  The WZNW Lagrangian for maps $g:\Sigma_2\to G$ from a two-dimensional
  Riemann surface $\Sigma_2$ to the supergroup $G$ is fully
  specified in terms of the invariant metric on $\g$ and it reads
\begin{equation}
  \label{eq:WZNWLagrangian}
  \mc{S}^{\text{WZNW}}[g]
  \ =\ -\frac{i}{4\pi}\int_{\Sigma_2}\langle g^{-1}\partial g,
  g^{-1}\bartial g\rangle\,dz\wedge d\bar{z}
       -\frac{i}{24\pi}\int_{B_3}\langle g^{-1}dg,[g^{-1}dg,g^{-1}dg]\rangle\ \ .
\end{equation}
  The second term is integrated over an auxiliary three-manifold
  $B_3$ which satisfies $\partial B_3=\Sigma_2$.
  Note that the measure $idz\wedge d\bar{z}$ is real. The
  topological ambiguity of the second term possibly imposes a
  quantization condition on the metric $\langle\cdot,\cdot\rangle$
  or, more precisely, on its bosonic restriction, in order to
  render the path integral well-defined.\footnote{Note that
  for WZNW models based on bosonic groups one usually explicitly
  introduces an integer valued constant, the level, which appears
  as a prefactor of the Killing form. For supergroups the Killing
  form might vanish. Hence there is no canonical normalization
  of the metric. Moreover, we would like to include models whose
  metric renormalizes non-multiplicatively (see below). Under
  these circumstances  it is not particularly convenient to
  display the level explicitly and we assume instead that all
  possible parameters are contained in the metric $\langle\cdot,
  \cdot\rangle$.} Given the parametrization \eqref{eq:Para}, the
  Lagrangian can be simplified considerably by making iterative
  use of the Polyakov-Wiegmann identity
\begin{equation}
  \label{eq:PW}
  \mc{S}^{\text{WZNW}}[gh]
  \ =\ \mc{S}^{\text{WZNW}}[g]+\mc{S}^{\text{WZNW}}[h]-\frac{i}{2\pi}\int
       \langle g^{-1}\bartial g,\partial hh^{-1}\rangle\,dz\wedge d\bar{z}\ \ .
\end{equation}
  The WZNW action evaluated on the individual fermionic bits vanishes
  because the invariant form \eqref{eq:Metric} is only supported
  on grade $0$ of the $\Integer$-grading. The final result is then
\begin{equation}
  \label{eq:WZNWLagrangianNew}
  \mc{S}^{\text{WZNW}}[g] \ = \ \mc{S}^{\text{WZNW}}[g_B,\theta]
  \ =\ \mc{S}^{\text{WZNW}}[g_B]
       -\frac{i}{2\pi}\int\langle\bartial\theta,g_B\,\partial\bar{\theta}
  \,g_B^{-1}\rangle\,dz\wedge d\bar{z}\ \ .
\end{equation}
  For the correct determination of the mixed bosonic and fermionic
  term it was again necessary to refer to the grading of
  $\g$. The latter implies for instance that the scalar product
  vanishes if bosonic generators are paired with fermionic ones.
\smallskip

  It is now crucial to realize (see also \cite{Schomerus:2005bf,Gotz:2006qp})
  that we may pass to an equivalent description of the WZNW model above by
  introducing an additional set of auxiliary fields $p_a$ and $\bar p^a$,
\begin{equation}
  \label{eq:FreeFieldInt}
  \begin{split}
    \mc{S}[g_B,p,\theta]
    &\ =\ \mc{S}_{\text{ren}}^{\text{WZNW}}[g_B]+\mc{S}_{\text{free}}
           [\theta,\bar{\theta},p,\bar{p}]
          +\mc{S}_{\text{int}}[g_B,p,\bar{p}]\\[2mm]
    &\ =\ \mc{S}_{\text{ren}}^{\text{WZNW}}[g_B]
          +\frac{i}{2\pi}\int\Bigl\{\langle p,\bartial\theta\rangle-
   \langle\bar{p},\partial\bar{\theta}\rangle-\langle p,g_B\,\bar{p}\,
   g_B^{-1}\rangle\Bigr\}\,dz\wedge d\bar{z}\ \ .
  \end{split}
\end{equation}
  Here, $\theta, \bar \theta$ and our new fermionic fields
  $p=p_aS_1^a$ and $\bar{p}=\bar{p}^aS_{2a}$ all take values in
  the Lie superalgebra $\g$. Our conventions may look slightly
  asymmetric but as we will see later this just resembles
  the asymmetry in the parametrization \eqref{eq:Para}.
  Up to certain subtleties that are encoded in the subscript
  ``ren'' of the
  first term, it is straightforward to see that we recover
  the original Lagrangian \eqref{eq:WZNWLagrangianNew} upon
  integrating out the auxiliary fields $p$ and $\bar{p}$.
\smallskip

  Let us comment a bit more on each term in the action
  \eqref{eq:FreeFieldInt}. Most importantly, we need to specify
  the renormalization of the bosonic WZNW model which results
  from the change in the path integral measure
  (cf.~\cite{Polyakov:1984et}). The computation of the relevant
  Jacobian has two important effects. First of all, it turns out
  that the construction of the purely bosonic WZNW model entering
  the action \eqref{eq:FreeFieldInt} employs the following
  renormalized metric\footnote{We assume this metric to
  be non-degenerate. Otherwise we would deal with what is known
  as the critical level or, in string terminology, the
  tensionless limit.}
\begin{equation}
  \label{eq:RenMetric}
  \langle K^i,K^j\rangle_{\text{ren}}\ =\ \kappa^{ij}-\gamma^{ij}
  \qquad\text{ with }\qquad
  \gamma^{ij}\ =\ \tr(R^iR^j)\ \ .
\end{equation}
  Note that this renormalization is not necessarily multiplicative.
  For simple Lie superalgebras the renormalized metric is always
  identical to the original one up to a factor. For non-simple
  Lie superalgebras, however, this is generically not the case as
  can be inferred from the example of $gl(1|1)$.
\smallskip

  As a second consequence of the renormalization, the action
  \eqref{eq:FreeFieldInt} may contain a Fradkin-Tseytlin term,
  coupling a non-trivial dilaton to the world-sheet curvature
  $R^{(2)}$,
\begin{equation}
  \label{eq:Dilaton}
  \mc{S}^{\text{WZNW}}_{\text{FT}}[g_b] \ = \
  \int_{\Sigma_2}d^2\sigma\sqrt{h}R^{(2)}\phi(g_B)\ \ \ \ \
  \text{where} \ \ \ \
  \phi(g_B)\ =\ -\frac{1}{2}\,\ln\det R(g_B)\ \ .
\end{equation}
  The same kind of expression has already been encountered in
  the investigation of the $GL(1|1)$ WZNW model, cf.\ %
  \cite{Rozansky:1992rx,Rozansky:1992td,Schomerus:2005bf}.
  From the discussion at the end of section \ref{ssc:KacMod} it is
  obvious that $\phi$ vanishes whenever $\g_0$ is a semisimple Lie
  algebra. Therefore, a non-trivial dilaton is a feature of the series
  $osp(2|2n)$, $sl(m|n)$ and $gl(m|n)$ or, in other words, of most basic
  Lie superalgebras of type~I. The precise reason for the claimed
  form of renormalization, i.e.\ the modification of the metric
  and the appearance of the dilaton, will become clear in the
  following sections when we discuss the full quantum symmetry
  of the supergroup WZNW model. At the moment let us just restrict
  ourselves to the comment that the dilaton is required in order
  to ensure the supergroup invariance of the path integral measure
  for the free fermion resolution, i.e.\ the description of the
  WZNW model in terms of the Lagrangian \eqref{eq:FreeFieldInt}.
\smallskip

  Before we conclude this subsection, let us quickly return to the
  fermionic terms of the Lagrangian \eqref{eq:FreeFieldInt} which
  may be rewritten in an even more explicit form using
\begin{equation}
  g_B\,\bar{p}\,g_B^{-1}
  \ =\ g_B\,S_{2b}\,\bar{p}^b\,g_B^{-1}
  \ =\ S_{2a}\,{R^a}_b(g_B)\,\bar{p}^b\ \ .
\end{equation}
  The result for the interaction term is
\begin{equation}
  \label{eq:FreeFieldIntTwo}
  \mc{S}_{\text{int}}[g_B,p,\bar p]\ =
   - \frac{i}{2\pi}\int p_a\,{R^a}_b(g_B)\,\bar{p}^b
   \,dz\wedge d\bar{z}\ \ .
\end{equation}
  In an operator formulation, the object ${R^a}_b(g_B)$ should be
  interpreted as a vertex operator of the bosonic WZNW model,
  transforming in the representation $R\otimes R^\ast$. We may
  consider the interaction term $p_a\,{R^a}_b(g_B)\,\bar{p}^b$
  as a screening current. Note that the latter is non-chiral by
  definition, a feature that is not really specific to supergroups
  but applies equally to bosonic models. Nevertheless, the
  existing literature on free field constructions did
  not pay much attention to this point. Actually, the distinction
  is not really relevant for purely bosonic WZNW models because of
  their simple factorization into left and right movers. In the
  present context, however, a complete non-chiral treatment
  must be enforced in order to capture and understand the
  special properties of supergroup WZNW models.

\subsection{\label{sc:Cov}Covariant formulation of the symmetry}

  It is well known that the full WZNW model exhibits a loop group symmetry.
  More precisely, the Lagrangian \eqref{eq:WZNWLagrangian} (and hence also
  the functional \eqref{eq:FreeFieldInt}) is invariant under
  multiplication of the field $g(z,\bar{z})$ with holomorphic elements
  from the left and with antiholomorphic elements from the right.
  Infinitesimally, each of these transformations generates an infinite dimensional
  current superalgebra, a central extension $\ag$ of the loop superalgebra
  belonging to $\g$. For the holomorphic sector the latter is equivalent to
  the following operator product expansions (OPEs). In the bosonic subsector
  we find
\begin{equation}
  \label{eq:OPEBB}
  K^i(z)\,K^j(w)\ =\ \frac{\kappa^{ij}}{(z-w)^2}+\frac{i{f^{ij}}_l\,K^l(w)}{z-w}\ \ .
\end{equation}
  The transformation properties of the fermionic currents are
\begin{align}
  \label{eq:OPEBF}
  K^i(z)\,S_1^a(w)&\ =\ -\frac{{(R^i)^a}_b\,S_1^b(w)}{z-w}&
  &\text{ and }&
  K^i(z)\,S_{2a}(w)&\ =\ \frac{{S_{2b}(w)\,(R^i)^b}_a}{z-w}\ \ .
\end{align}
  Finally we need to specify the OPE of the fermionic currents,
\begin{equation}
  \label{eq:OPEFF}
  S_1^a(z)\,S_{2b}(w)\ =\ \frac{\delta_b^a}{(z-w)^2}-\frac{{(R^i)^a}_b\,
  \kappa_{ij}\,K^j(w)}{z-w}\ \ .
\end{equation}
  The previous operator product expansions are straightforward
  extensions of the commutation relations \eqref{eq:CommBB},
  \eqref{eq:CommBF} and \eqref{eq:CommFF}. The central extension
  is determined by the invariant metric \eqref{eq:Metric}.%
\smallskip

  The current superalgebra above defines a chiral vertex algebra
  via the Sugawara construction \cite{Sugawara:1968rw}. As usual,
  the corresponding energy momentum tensor is obtained by contracting
  the currents with the inverse of a distinguished invariant and
  non-degenerate metric. The appropriate fully renormalized (hence
  the subscript ``full-ren'') metric is defined by
\begin{equation}
  \label{eq:FulMetric}
  \begin{split}
    \langle K^i,K^j\rangle_{\text{full-ren}}
    &\ =\ (\Omega^{-1})^{ij}\ =\ \kappa^{ij}-\gamma^{ij}-\frac{1}{2}\,{f^{im}}_n\,{f^{jn}}_m\\[2mm]
    \langle S_1^a,S_{2b}\rangle_{\text{full-ren}}
    &\ =\ {(\Omega^{-1})^a}_b\ =\ \delta_b^a+{(R^i\kappa_{ij}R^j)^a}_b
  \end{split}
\end{equation}
  and it is the result of adding half the Killing form of the Lie superalgebra
  $\g$ to the original classical metric \eqref{eq:Metric}.\footnote{Again, this
  renormalization does not need to be multiplicative, see for instance $GL(1|1)$.}
  Note that some of the terms in the fully renormalized metric \eqref{eq:FulMetric}
  can be identified with the (partially) renormalized metric \eqref{eq:RenMetric}
  which we introduced while deriving the free fermion Lagrangian. The energy
  momentum tensor of our theory involves the inverse of the fully renormalized
  metric,
\begin{equation}
  \label{eq:AffineT}
  T\ =\ \frac{1}{2}\,\Bigl[K^i\,\Omega_{ij}\,K^j-S_1^b{\Omega^a}_bS_{2a}+
  S_{2a}{\Omega^a}_bS_1^b\bigr]\ \ .
\end{equation}
  Both, currents and energy momentum tensor, may similarly be defined
  for the antiholomorphic sector. The appearance of a renormalized metric
  in the Sugawara construction is a rather common feature. Supergroup WZNW
  models are certainly not exceptional in this respect.
\smallskip

  In order to complete the discussion of the operator content, we have to
  introduce vertex operators $\Phi^{(\mc{M})}(z,\bar{z})$. The latter
  carry a representation $\mc{M}$ of $\g\oplus\g$, the underlying horizontal
  part of the current superalgebra of our model. If we assume for a moment that
  $\mc{M}=(\mu\nu)$ where $\mu$ and $\nu$ refer to Kac modules of
  the individual factors in $\g\oplus\g$ then primary fields are
  characterized by the operator products
\begin{align}
  \label{eq:VertexBos}
    K^i(z)\,\Phi^{(\mu\nu)}(w,\bar{w})
    &\ =\ -\frac{D^{(\mu)}(K^i)\Phi^{(\mu\nu)}(w,\bar{w})}{z-w}&
    S_{2a}(z)\,\Phi^{(\mu\nu)}(w,\bar{w})
    &\ =\ 0\\[2mm]
    \bar{K}^i(\bar{z})\,\Phi^{(\mu\nu)}(w,\bar{w})
    &\ =\ \frac{\Phi^{(\mu\nu)}(w,\bar{w})D^{(\nu)}(K^i)}{\bar{z}-\bar{w}}&
    \bar{S}_1^a(\bar{z})\,\Phi^{(\mu\nu)}(w,\bar{w})
    &\ =\ 0\ \ .
\end{align}
  In addition, there are fields
  $(S_1^{a_1}\cdots S_1^{a_s}\bar{S}_{2b_1}\cdots\bar{S}_{2b_t}\Phi^{(\mu\nu)})(z,\bar{z})$
  which belong to the same representation of the horizontal
  subsuperalgebra. The matrices $D^{(\mu)}$ are representation matrices
  of $\g_0$. As usual we may infer the conformal dimension of the primary
  fields from their operator product expansion with the energy momentum
  tensor,
\begin{equation}
  \begin{split}
    T(z)\,\Phi^{(\mu\nu)}(w,\bar{w})
    &\ =\ \frac{h^{(\mu\nu)}\,\Phi^{(\mu\nu)}(w,\bar{w})}{(z-w)^2}+
  \frac{\partial\Phi^{(\mu\nu)}(w,\bar{w})}{z-w}\\[2mm]
    \bar{T}(\bar{z})\,\Phi^{(\mu\nu)}(w,\bar{w})
    &\ =\ \frac{\bar{h}^{(\mu\nu)}\,\Phi^{(\mu\nu)}(w,\bar{w})}{(z-w)^2}+
  \frac{\bartial\Phi^{(\mu\nu)}(w,\bar{w})}{\bar{z}-\bar{w}}\ \ .
  \end{split}
\end{equation}
  Using the standard techniques one easily finds that the
  conformal dimensions are given by (renormalized) Casimir
  eigenvalues,
\begin{align}
  h^{(\mu\nu)}
  &\ =\ \frac{1}{2}\,C_\mu^{\text{full-ren}}&
  \bar{h}^{(\mu\nu)}
  &\ =\ \frac{1}{2}\,C_\nu^{\text{full-ren}}\ \ .
\end{align}
  The corresponding Casimir is given by
  $C^{\text{full-ren}}=K^i\Omega_{ij}K^j+\tr(\Omega R^i)\kappa_{ij}K^j$
  and should be thought of as a renormalization of
  eq.\ \eqref{eq:Eigenval}. It is
  important to stress once more that in our conventions
  the level is contained implicitly in the metric
  $\kappa^{ij}$. Thus the conformal dimensions depend
  on the level. They vanish if the metric of the supergroup
  is scaled to infinity. In that limit the ground state sector decouples,
  and it can be analyzed using methods of harmonic analysis.
  This will be carried out in section \ref{sc:SCA}.

\subsection{\label{sc:AlgFF}Free fermion resolution}

  Our next aim is to describe the current superalgebra defined above
  and the associated primary fields in terms of the decoupled system
  of bosons and fermions that appear in the Lagrangian
  \eqref{eq:FreeFieldInt}. As one of our ingredients we shall
  employ the bosonic current algebra
\begin{equation}
  K_B^i(z)\,K_B^j(w)\ =\ \frac{(\kappa-\gamma)^{ij}}{(z-w)^2}+
  \frac{i{f^{ij}}_l\,K_B^l(w)}{z-w}\ \ ,
\end{equation}
  which is defined using the (partially) renormalized metric
  which has been introduced in \eqref{eq:RenMetric}. In addition, we
  need $r$ free fermionic ghost systems with fields $p_a(z)$ and
  $\theta^a(z)$ of spins $h=1$ and $h=0$, respectively. They
  possess the usual operator products
\begin{equation}
  p_a(z)\,\theta^b(w)\ =\ \frac{\delta_a^b}{z-w}\ \ .
\end{equation}
  Fermionic fields are assumed to have trivial operator product
  expansions with the bosonic generators. By construction, the
  currents $K_B^i$ and the fields $p_a$, $\theta^b$ generate the
  chiral symmetry of the field theory whose action is
\begin{equation}
  \label{eq:WZNWDecoupled}
  \mc{S}_0[g_B,p,\theta]
  \ =\ \mc{S}_{\text{ren}}^{\text{WZNW}}[g_B]+\mc{S}_{\text{free}}
  [\theta,\bar{\theta},p,\bar{p}]\ \ .
\end{equation}
  Our full WZNW theory may be considered as a deformation of this
  theory, once we take into account the interaction term between bosons
  and fermions, see eq.\ \eqref{eq:FreeFieldIntTwo}. The further development
  of this approach and its consequences will be the subject of
  section \ref{sc:Quantum}.
\smallskip

  But returning first to the decoupled action \eqref{eq:WZNWDecoupled},
  it is easy to see that it defines a conformal field theory with energy
  momentum tensor
\begin{equation}
  \label{eq:FreeT}
  T\ =\ \frac{1}{2}\,\Bigl[K_B^i\,\Omega_{ij}\,K_B^j+
  \tr(\Omega R^i)\,\kappa_{ij}\,\partial K_B^j\Bigr]
        -p_a\partial\theta^a\ \ .
\end{equation}
  Note the existence of the dilaton contributions, i.e.\ terms
  linear in derivatives of the currents. In addition to the conformal
  symmetries, the action \eqref{eq:WZNWDecoupled} is also invariant
  under a $\ag\oplus\ag$ current superalgebra. The corresponding
  holomorphic currents are defined by the relations (normal ordering
  is implied)
\begin{equation}
  \label{eq:FFRes}
  \begin{split}
    K^i(z)&\ =\ K_B^i(z)+p_a\,{(R^i)^a}_b\,\theta^b(z)\\[2mm]
    S_1^a(z)&\ =\ \partial\theta^a(z)+{(R^i)^a}_b\,\kappa_{ij}\,
    \theta^bK_B^j(z)-\frac{1}{2}{(R^i)^a}_c\,\kappa_{ij}\,
    {(R^j)^b}_d\,p_b\theta^c\theta^d(z)\\[2mm]
    S_{2a}(z)&\ =\ -p_a(z)\ \ .
  \end{split}
\end{equation}
  It is a straightforward exercise, even though slightly
  cumbersome and lengthy, to check that this set of
  generators reproduces the operator product expansions
  \eqref{eq:OPEBB}, \eqref{eq:OPEBF} and \eqref{eq:OPEFF}.
  The only input we need is the Jacobi identity \eqref{eq:Jacobi}.
  The same identity shows that the quantity in \eqref{eq:FFRes}
  which is used to contract $p_b\theta^c\theta^d$ is in fact
  antisymmetric in the lower two indices. Obviously, a similar set
  of currents may be obtained for the antiholomorphic sector.
  Given the representation \eqref{eq:FFRes} for the current
  superalgebra one may also check the equivalence of the
  expressions \eqref{eq:AffineT} and \eqref{eq:FreeT} for the
  energy momentum tensors. Algebraically, the calculation rests
  on the Jacobi identity \eqref{eq:Jacobi} as well as on the
  equations
\begin{equation}
  (\Omega^{-1})^{ij}\,\kappa_{ij}\,{(R^l)^a}_b
  \ =\ {(R^i)^a}_c\,{(\Omega^{-1})^c}_b
  \ =\ {(\Omega^{-1})^a}_c\,{(R^i)^c}_b\ \ .
\end{equation}
  The latter arise as invariance constraints for the metric
  $\langle\cdot,\cdot\rangle_{\text{full-ren}}$ as defined
  in eq.\ \eqref{eq:FulMetric}.
\smallskip

  The current superalgebra defined in \eqref{eq:FFRes} has a natural
  action on the vertex operators of the conformal field theory
  defined by the decoupled Lagrangian $\mc{S}_0$. Once we include
  the interaction term, the theory becomes equivalent to the
  full WZNW model. Hence, we must be able to map the vertex
  operators of the decoupled theory to the vertex operators of
  the WZNW theory. The precise relation turns out to be rather
  involved. Therefore, we postpone a more detailed exposition of
  this relation to section \ref{sc:Quantum}. Instead, we will
  continue with a semi-classical analysis of the space of vertex
  operators. This procedure allows us to clearly exhibit
  the subtleties of the full quantum field theory in a simple
  and geometric setup.

\section{\label{sc:SCA}Semi-classical analysis}

  The WZNW model we introduced in the last section admits a semi-classical
  limit when the invariant metric defined in \eqref{eq:Metric} is scaled
  to infinity. This corresponds to choosing the levels of the underlying
  bosonic WZNW model large. In this weak curvature regime we expect the
  conformal dimensions of all primary fields to tend to zero and the
  higher modes to decouple. We will start with a discussion of the
  global symmetry of the WZNW model and how it is realized in terms of
  differential operators on the space of quantum mechanical wave functions.
  Then we discuss the Laplacian, i.e.\ the wave operator, on $G$
  and determine its (generalized) eigenfunctions and eigenvalues which
  approximate the vertex operators and their conformal
  dimensions in the full conformal field theory. It is shown that the
  spectrum contains non-chiral indecomposable modules on which the
  Laplacian is not diagonalizable.

\subsection{Symmetry}

  One of the inherent properties of supergroup manifolds $G$ is that they
  admit two actions of $G$ on itself. These so-called left and right
  regular actions are defined by the maps
\begin{equation}
  \label{eq:LRaction}
  L_h:\quad g\ \mapsto\ hg
  \qquad\text{ and }\qquad
  R_h:\quad g\ \mapsto\ gh^{-1}\ \ .
\end{equation}
  Since the definition of the WZNW Lagrangian \eqref{eq:WZNWLagrangian} only
  involves the invariant metric, both actions are automatically symmetries of
  our model. In fact, in the present situation they are even promoted to
  current superalgebra symmetries as we have already seen in the previous
  section. In this section we will just discuss the
  point-particle limit (or minisuperspace approximation) where only the
  zero-modes are taken into account and every dependence on the world-sheet
  coordinates is ignored. This corresponds to quantum mechanics on the supergroup
  \cite{Gepner:1986wi}. Our aim is to find all the eigenfunctions of
  the Laplace (or wave) operator.
\smallskip

  Given the symmetry above we know that the state space of the physical
  system may be decomposed into representations of $\g\oplus\g$.
  The corresponding symmetry can be realized in terms of differential
  operators acting on the wave functions which are elements of some
  algebra of functions $\mc{F}(G)$ on the supergroup.\footnote{The naive
  definition of the algebra of function as elements of the Grassmann
  algebra in the fermions $\theta^a$ and $\bar{\theta}_a$ with square
  integrable coefficient functions on $G_B$ leads to inconsistencies. A more
  detailed discussion of these subtle points and the explicit introduction
  of the correct algebra of function shall be postponed until section
  \ref{sc:SpectrumHA}.}
  These functions will naturally depend on a bosonic group element $g_B$ and on the
  fermionic coordinates $\theta^a$ and $\bar{\theta}_a$. By using a
  Taylor expansion with respect to the fermionic variables the basis
  elements of $\mc{F}(G)$ may be represented as a complex valued
  function depending solely on $g_B$ multiplied by a product of
  Grassmann variables.
\smallskip

  The left and right regular action
  of the supergroup on itself, as given in \eqref{eq:LRaction},
  then induces the action
\begin{equation}
  (h_L\times h_R)\cdot f:\quad g\ \mapsto\ f(h_L^{-1}gh_R)
\end{equation}
  on arbitrary elements $f\in\mc{F}(G)$. This in turn translates into the following
  differential operators,
\begin{equation}
  \label{eq:DiffL}
  \begin{split}
    K^i&\ =\ K_B^i-{(R^i)^a}_b\,\theta^b\,\partial_a
    \hspace{3cm}S_{2a}\ =\ -\partial_a\\[2mm]
    S_1^a&\ =\ {R^a}_b(g_B)\,\bartial^b+{(R^i)^a}_b\,\theta^b\,
   \kappa_{ij}\,K_B^j-\frac{1}{2}\,{(R^i)^a}_c\,\kappa_{ij}\,
   {(R^j)^b}_d\,\theta^c\theta^d\partial_b\ \ ,
  \end{split}
\end{equation}
  for the infinitesimal left regular action. In addition to the various structure
  constants of the Lie superalgebra, these expressions contain derivatives $\partial_a
  =\partial/\partial\theta^a$ and $\bartial^a=\partial/\partial \bar{\theta}_a$ with
  respect to the Grassman variables $\theta^a$ and $\bar{\theta}_a$. We have also
  introduced the differential operators $K_B^i$ which implement the regular action
  of the bosonic subgroup $G_B$. They involve derivatives with respect
  to bosonic coordinates only, but the precise form depends on the
  particular choice of coordinates on $G_B$. Similar expressions
  can be found for the infinitesimal generators of the
  right action,
\begin{equation}
  \label{eq:DiffR}
  \begin{split}
    \bar{K}^i&\ =\ \bar{K}_B^i+\bar{\theta}_a\,{(R^i)^a}_b\,\bartial^b
    \hspace{3cm}\bar{S}_1^a\ =\ \bartial^a\\[2mm]
    \bar{S}_{2a}&\ =\ -{R^b}_a(g_B)\,\partial_b-\bar{\theta}_b\,{(R^i)^b}_a\,
    \kappa_{ij}\,\bar{K}_B^j
                      -\frac{1}{2}\,{(R^i)^c}_a\,\kappa_{ij}\,{(R^j)^d}_b\,
    \bar{\theta}_c\bar{\theta}_d\bartial^b\ \ .
  \end{split}
\end{equation}
  One can check explicitly that these two sets of differential operators
  form two (anti)commuting copies of the Lie superalgebra $\g$.
  Again, these calculations rely heavily on the Jacobi identity
  \eqref{eq:Jacobi}.%
\smallskip

  The expressions for the differential operators exhibit some peculiar
  properties that we would like to expand on. Note that, apart from purely
  bosonic pieces, the generators \eqref{eq:DiffL} of the left regular action would
  only involve the Grassmann coordinates $\theta_a$ and the corresponding
  derivatives -- but no bared fermions -- if it were not for the very first term
  in the definition of $S_1^a$. Indeed, this term does contain derivatives
  with respect to the fermionic coordinates $\bar{\theta}_a$. Obviously, the
  situation is reversed for the right regular action. It is also
  worth stressing that the coefficients in the first terms of both
  $S_1^a$ and $\bar{S}_{2a}$ are non-trivial functions on the bosonic
  group. Again this is in sharp contrast to all the other terms whose
  coefficients are independent of the bosonic coordinates (though
  functions of the Grassmann variables, of course). It has been
  emphasized in \cite{Gotz:2006qp} that the occurrence of the matrix
  $R(g_B)$ can spoil the normalizability properties of the functions
  the symmetry transformations are acting on. This always happens if
  the target space is non-compact since $R$ is a finite dimensional
  representation and hence non-unitary in that case. Consequently, the
  product of an $L^2$-function from $\mc{F}(G_B)$ with $R(g_B)$ will
  not be an $L^2$-function anymore.
\smallskip

  In view of these issues with $S_1^a$ and $\bar{S}_{2a}$ it is tempting
  to simply drop the troublesome terms. Even though that might seem a rather
  arbitrary modification at first, it turns out that the corresponding
  truncated differential operators $\mb{K}^i = K^i$, $\mb{S}_{2a}=S_{2a}$,
\begin{equation}
  \mb{S}_1^a \ = \ {(R^i)^a}_b\,\theta^b\,\kappa_{ij}\,K_B^j-
  \frac{1}{2}\, {(R^i)^a}_c\,\kappa_{ij}\,{(R^j)^b}_d\,\theta^c
  \theta^d\partial_b
\end{equation}
  and their bared analogues also satisfy the commutation relations
  of $\g\oplus\g$! For the special case of $PSU(1,1|2)$, it was
  explained in \cite{Gotz:2006qp} that this is much more than a
  mere curiosity. Indeed, we conclude that the truncated operators 
  $\mb{K}^i$, $\mb{S}_1^a$ and $\mb{S}_{2a}$ model
  the action of zero-modes of our currents \eqref{eq:FFRes} on ground
  states in the decoupled free fermion theory, i.e.\ before the
  coupling of bosonic and fermionic fields is taken into account.
  Note that the zero-mode of $p(z)$ is a field theoretic
  incarnation of the derivative $\partial$ since $p(z)$ is the
  canonically conjugate momentum belonging to $\theta(z)$. We
  shall now proceed to argue that the original differential
  operators \eqref{eq:DiffL} and \eqref{eq:DiffR} encode a much
  more intricate structure, namely the action of the zero-modes
  on primaries in the full interacting WZNW model.

\subsection{\label{sc:HarmAn}Harmonic analysis}

  The algebra of functions $\mc{F}(G)$ furnishes a representation
  of $\g\oplus\g$ via the differential operators \eqref{eq:DiffL} and
  \eqref{eq:DiffR}. Our aim is to write $\mc{F}(G)$ as a direct sum
  of indecomposable building blocks of the type discussed in section
  \ref{sc:RepTh}. The final result can be found in eq.\ \eqref{eq:SpectrumFull}
  below. But since the outcome is rather complicated and somewhat
  hard to digest we would like to start the harmonic analysis by
  discussing the left and the right action of $\g$ separately. We
  claim that the space of functions decomposes under these actions
  according to\footnote{A similar expression already appeared in
  \cite{Huffmann:1994ah} in a more general context.}
\begin{equation}
  \label{eq:SpectrumLR}
  \mc{F}(G)\bigr|_{\g(\text{left})}
  \ =\ \mc{F}(G)\bigr|_{\g(\text{right})}
  \ =\ \bigoplus_{\mu\in\Typ(G)}\,\dim(\mc{K}_\mu)\;\mc{K}_\mu
       \oplus\bigoplus_{\mu\in\At(G)}\,\dim(\mc{L}_\mu)\;\mc{P}_\mu\ \ .
\end{equation}
  The symbols $\Typ(G)$ and $\At(G)$ denote the sets of typical
  and atypical irreducible representations of the supergroup. The
  distinction between modules of $G$ and modules of $\g$ is necessary
  since there might exist representations of the Lie superalgebra which
  cannot be lifted to $G$. Under rather general conditions (to be
  recalled below eq.~\eqref{eq:PeterWeyl}) the set $\Rep(G)$ of
  supergroup representations coincides with
  $\Rep(G_B)\subset\Rep(\g_0)$, the set of all unitary irreducible
  representations of the bosonic subgroup $G_B$.
\smallskip

  As we see, the decomposition \eqref{eq:SpectrumLR} clearly
  distinguishes between the
  typical and the atypical sector of our space. In the typical
  sector we sum over irreducible Kac modules $\mc{K}_\mu=\mc{L}_\mu$
  with a multiplicity space $\mc{M}(\mc{K}_\mu)$ of dimension
  $\dim\mc{K}_\mu$, a prescription which is familiar from the
  Peter-Weyl theory for bosonic groups. In contrast, the atypical
  sector consists of a sum over all the projective covers $\mc{P}_\mu$
  belonging to atypical irreducibles $\mc{L}_\mu$ and coming with a
  multiplicity space $\mc{M}(\mc{P}_\mu)$ of the smaller dimension
  $\dim\mc{L}_\mu<\dim\mc{K}_\mu$. Note that the algebra of functions
  forms a projective module and hence possesses a Kac composition
  series, i.e.\ a filtration in terms of Kac modules. This immediately
  permits us to spell out the character of the $\g\oplus\g$-module
  $\mc{F}(G)$ and it will lead to a concrete proposal for the modular
  invariant partition function of the WZNW model in section
  \ref{sc:Quantum}.
\smallskip

  Naturally, our formula \eqref{eq:SpectrumLR} is the same for the
  left and the right action. This symmetry between left and right
  regular transformations must certainly be maintained when we
  extend our analysis to the combined left and right action of
  $\g \oplus \g$ on $\mc{F}(G)$. In the typical sector the
  multiplicity spaces of the Kac modules have precisely the
  dimension that is needed to promote them to Kac modules
  themselves, a prescription that is perfectly consistent
  with the symmetry between left and right action. On the other
  hand, the same symmetry requirement excludes that the individual
  multiplicity spaces in the atypical sector are simply
  promoted to irreducible representations of $\g$.
  Consequently, the left action must induce maps between
  various multiplicity spaces for the right action and vice
  versa. In this way, the atypical sector then consists of
  non-chiral indecomposables $\mc{I}_{[\sigma]}$ which entangle
  a (possibly infinite) number of left and right projective covers
  whose labels belong to the same block $[\sigma]$. The final
  expression for the representation content of the algebra of
  functions on $G$ is thus
  of the form
\begin{equation}
  \label{eq:SpectrumFull}
  \begin{split}
    \mc{F}(G)\bigr|_{\g\oplus\g}
    &\ =\ \bigoplus_{\mu\in\Typ(G)}\,\mc{L}_\mu\otimes\mc{L}_\mu^\ast
          \oplus\bigoplus_{[\sigma]\in\Gamma_{\text{atyp}}(G)}\,\mc{I}_{[\sigma]}\ \ .
  \end{split}
\end{equation}
  The systematic study of the non-chiral representations
  $\mc{I}_{[\sigma]}$ will be left for future work. Note that
  similar and, in the specific cases of $GL(1|1)$ and
  $SU(2|1)$, more explicit expressions have been obtained in
  \cite{Schomerus:2005bf,Gotz:2006qp,Saleur:2006tf}. We also
  wish to emphasize that the socle of \eqref{eq:SpectrumFull},
  i.e.\ its maximal semisimple subspace, corresponds to a direct
  sum over all pairs of irreducible representations and their
  duals. It would be interesting to compare our findings with
  the more abstract results in \cite{Scheunert2002:MR1927431}
  where the space of functions on $G=GL(m|n)$ is treated in
  the framework of Hopf superalgebras.
\smallskip

  Having stated the main results of this subsection we would like to
  sketch their derivation. For the proof of eq.~\eqref{eq:SpectrumLR},
  it is advantageous to enlarge the symmetry from $\g$ to an
  action $\g \oplus \g_0$, i.e.\ to retain the bosonic generators
  of the right regular transformations if we analyze the left
  action. With respect to the combined action one finds
\begin{align}
  \label{eq:SpectrumMixed}
  \mc{F}(G)\bigr|_{\g\oplus\g_0}
  &\ =\ \bigoplus_{\mu\in\Rep(G_B)}\,\mc{B}_\mu\otimes V_\mu^\ast&
  \mc{F}(G)\bigr|_{\g_0\oplus\g}
  &\ =\ \bigoplus_{\mu\in\Rep(G_B)}\,V_\mu\otimes\mc{B}_\mu^\ast\ \ .
\end{align}
  In fact, from the Peter-Weyl theorem for compact semisimple Lie
  groups (or suitable generalizations thereof) we deduce that
  the functions
\begin{equation}
  \label{eq:CandFun}
  \det R(g_B^{-1})\,{\bigl[D^{(\mu)}(g_B)\bigr]^\alpha}_\beta\,
  \theta^1\cdots\theta^r\,\bar{\theta}_1\cdots\bar{\theta}_r
\end{equation}
  involving matrix elements of the representation $D^{(\mu)}$ are
  part of the spectrum for all unitary irreducible representations
  $\mu$ of $G_B$. The matrix elements of $D^{(\mu)}$ transform in the
  representation $V_\mu\otimes V_\mu^\ast$ with respect to $\g_0
  \oplus\g_0$. Since the product of the remaining factors multiplying
  $D^{(\mu)}$ is invariant under purely bosonic transformations, we
  conclude that the set of functions \eqref{eq:CandFun} transforms in
  $V_\mu\otimes V_\mu^\ast$ as well.
\smallskip

  All that remains to be done is to augment the action on the left from
  the bosonic subalgebra $\g_0$ to the entire Lie superalgebra
  $\g$. The supersymmetric multiplets we generate from the
  functions \eqref{eq:CandFun} by repeated action with all the
  fermionic generators $S_1^a$ and $S_{2a}$ are isomorphic to
  the representation $\mc{B}_\mu$ of $\g$. Similar remarks apply
  if we consider the action of $\g_0\oplus\g$. Thereby we have
  established the decompositions \eqref{eq:SpectrumMixed}. In
  order to proceed from eqs.\ \eqref{eq:SpectrumMixed} to the
  decomposition formulas \eqref{eq:SpectrumLR} the representations
  $\mc{B}_\mu$ must be decomposed into their indecomposable building
  blocks. This is achieved  with the help of
  eq.~\eqref{eq:BStructure} and results in eq.\ \eqref{eq:SpectrumLR}
  after a simple re-summation. Our derivation has actually
  furnished a slightly stronger result since it determines
  how the multiplicity spaces decompose with respect to
  the action of the bosonic subalgebra $\g_0$.

\subsection{\label{sc:SpectrumHA}Spectrum and generalized eigenfunctions}

  Given the decomposition of the algebra of functions into representations
  of $\g\oplus\g$ we can now address our original problem of finding the
  semi-classical expressions of both the conformal dimensions and
  the primary fields. In the semi-classical limit, conformal dimensions
  are given by (half) the eigenvalues of the Casimir operator acting on
  $\mc{F}(G)$. Since we are dealing with a space of functions we will
  refer to the latter as ``Laplacian'' on the supergroup. The eigenvalues
  can be read off directly from the decomposition \eqref{eq:SpectrumFull}.
  In the typical sector the Laplacian is diagonalizable and leads to the
  eigenvalues $C(\mc{K}_\mu)$. On the other hand, the Laplacian ceases to
  be diagonalizable on the non-chiral representations $\mc{I}_{[\sigma]}$.
  Here, the Casimir may just be brought into Jordan normal form.
\smallskip

  The previous paragraph provides a complete solution of the eigenvalue
  problem but it does not yield explicit formulas for the (generalized)
  eigenfunctions. Since the latter are semi-classical versions of the
  primary fields in the full CFT (see section \ref{sc:Quantum} below), it
  seems worthwhile recalling the elegant construction of eigenfunctions
  that was presented recently in \cite{Gotz:2006qp}. The Laplace operator
  on our supergroup $G$ is given by
\begin{equation}
  \label{eq:Laplace}
  \Delta
  \ =\ \frac{1}{2}\,C
  \ =\ \Delta_B-\frac{1}{2}\tr(R^i)\,\kappa_{ij}\,K_B^j-\,\partial_a\,{R^a}_b(g_B)\,\bartial^b\ \ .
\end{equation}
  Observe that only the last term contains fermionic derivatives, with
  coefficents which depend on bosonic coordinates. Let us also emphasize
  that the purely bosonic piece of $\Delta$ differs from the Laplacian on
  the bosonic subgroup by the second term. This deviation is related to
  the presence of the non-trivial dilaton contribution \eqref{eq:Dilaton}.
  Since the complete Laplacian is non-diagonalizable it was
  proposed in \cite{Gotz:2006qp} to perform the harmonic analysis
  in two steps. First an auxiliary problem is solved which is based
  on the purely bosonic Laplacian
\begin{equation}
  \Delta_0\ =\ \Delta_B-\frac{1}{2}\,\tr(R^i)\,\kappa_{ij}\,K_B^j\ \ .
\end{equation}
  This auxiliary Laplacian agrees with the Casimir operator obtained
  from the reduced differential operators $\mb{K}$ and $\mb{S}$ and,
  as we shall see, it is completely diagonalizable on the following
  auxiliary space\footnote{The auxiliary space $\mb{F}(G)$ should be
  thought of as the semi-classical truncation of the state space for
  the decoupled theory $\mc{S}_0$, see eq.~\eqref{eq:WZNWDecoupled}.
  On the other hand $\mc{F}(G)$ corresponds to the semi-classical
  truncation of the full state space of the WZNW model.}
\begin{equation}
  \label{eq:Falg}
  \mb{F}(G)\ =\ \mc{F}(G_B)\otimes\bigwedge(\theta^a,\bar{\theta}_b)\ \ .
\end{equation}
  Here, the factor $\mc{F}(G_B)$ denotes the algebra of square
  (or $\delta$-function) normalizable functions on the bosonic subgroup
  and $\bigwedge(\theta^a,\bar{\theta}_b)$ is the Grassmann (or exterior)
  algebra generated by the fermionic coordinates. In the second step, the
  eigenfunctions of $\Delta_0$ are mapped to generalized eigenfunctions of
  $\Delta$ using a linear map $\Xi:\mb{F}(G)\to\mc{F}(G)$. The
  latter adds ``subleading'' fermionic contributions in a formal but
  well-defined way and thereby turns an eigenfunction of $\Delta_0$ into
  a generalized eigenfunction of $\Delta$. Our prescription involves
  explicit multiplications with the matrix elements of $R(g_B)$ which,
  e.g.\ for non-compact groups $G_B$, are not necessarily part of the
  unitary spectrum. Hence, the eigenfunctions of $\Delta$ need not be
  normalizable in the original sense, i.e.\ when regarded as Grassmann
  valued functions on the bosonic subgroup. This is the main reason why
  we need to distinguish between the spaces $\mb{F}(G)$ and
  $\mc{F}(G)=\text{Im}(\Xi)$. Ultimately, the problem may be
  traced back to the presence of the terms involving $R(g_B)$ in
  $S_1^a$ and $\bar{S}_{2a}$. In fact, as we pointed out before,
  because of those terms the unreduced differential operators
  may cease to act within $\mb{F}(G)$.
\smallskip

  In order to gain some intuition into the structure of the
  function space \eqref{eq:Falg} as a representation
  of the symmetry algebra $\g\oplus\g$, it is helpful
  to restrict the action to the bosonic subalgebra
  $\g_0\oplus\g_0$ first. Since the differential operators
  $\mb{K}^i$ and $\bar{\mb{K}}^i$ factorize in an action on the
  function algebra $\mc{F}(G_B)$ and on the Grassmann algebra
  $\bigwedge(\theta^a,\bar{\theta}_b)$, we can decompose
  both factors separately. If the bosonic subgroup is
  compact, semisimple and simply-connected we may employ
  the Peter-Weyl theorem in order to obtain
\begin{equation}
  \label{eq:PeterWeyl}
  \mc{F}(G_B)\bigr|_{\g_0\oplus\g_0}
  \ =\ \bigoplus_{\mu\in\Rep(G_B)}\,V_\mu\otimes V_\mu^\ast\ \ ,
\end{equation}
  where $\Rep(G_B)\subset\Rep(\g_0)$ is the set of all unitary irreducible
  representations of $G_B$. In more general situations this
  formula will need a slight refinement concerning the content of
  $\Rep(G_B)$, although the structure will still be very similar.
  With regard to the fermions, the left action just affects the
  set $\theta^a$, while the right action operates on the set
  $\bar{\theta}_a$. Given the known transformation behavior
  of a single fermion we thus find
\begin{equation}
  \bigwedge(\theta^a,\bar{\theta}_b)\bigr|_{\g_0\oplus\g_0}
  \ =\ \mc{F}\otimes\mc{F}^\ast\ \ .
\end{equation}
  Combining these simple facts and defining $\Rep(G)=\Rep(G_B)$
  we conclude
\begin{equation}
  \label{eq:FBosDeco}
  \mb{F}(G)\bigr|_{\g_0\oplus\g_0}
  \ =\ \bigoplus_{\mu\in\Rep(G)}\Bigl[V_\mu\otimes\mc{F}\Bigr]
  \otimes\Bigl[V_\mu\otimes\mc{F}\Bigr]^\ast\ \ .
\end{equation}
  Before we proceed to the supersymmetric extension, we would like to
  discuss the general form of elements in the individual subspaces
  of \eqref{eq:FBosDeco}. The space of functions is spanned by
\begin{equation}
  f_{b_1\cdots b_t,\beta}^{(\mu)a_1\cdots a_s,\alpha}(g)
  \ =\ {\bigl[D^{(\mu)}(g_B)\bigr]^\alpha}_\beta\ \theta^{a_1}\cdots\theta^{a_s}
       \,\bar{\theta}_{b_1}\cdots\bar{\theta}_{b_t}\ \ ,
\end{equation}
  where $D^{(\mu)}$ denotes the representation of the bosonic subgroup $G_B$
  on the module $V_\mu$ .
\smallskip

  Our most important task is to determine how the bosonic representations
  that occur in the decomposition \eqref{eq:FBosDeco} combine into multiplets
  of the full symmetry $\g\oplus\g$. As a first hint on what the answer will
  be, we observe that the representation content in eq.\ \eqref{eq:FBosDeco}
  agrees with the bosonic content of Kac modules. And indeed, under the
  action of fermionic generators, the various bosonic modules are
  easily seen to combine into our modules $\mc{K}_\mu$. To see
  this we note that the purely bosonic functions ${\bigl[D^{(\mu)}(g_B)
  \bigr]^\alpha}_\beta$ are annihilated by $\mb{S}_{2a}$ and
  $\bar{\mb{S}}_1^a$ simultaneously and therefore they span the
  subspace $V_\mu\otimes V_\mu^\ast$ from which we induce
  the Kac module $\mc{K}_\mu\otimes\mc{K}_\mu^\ast$. Consequently,
  we obtain the decomposition
\begin{equation}
  \mb{F}(G)\bigr|_{\g\oplus\g}
  \ =\ \bigoplus_{\mu\in\Rep(G)}\mc{K}_\mu\otimes\mc{K}_\mu^\ast\ \ .
\end{equation}
  Note that the sum runs over both typical and atypical
  representations, i.e.\ the space of functions is not fully
  reducible. The Laplacian $\Delta_0$ is completely
  diagonalizable on this space and its eigenvalues are given
  by eq.~\eqref{eq:Eigenval}.
\smallskip

  Let us now return to the analysis of the space $\mc{F}(G)$.
  We recall that a function $\Phi_\lambda\in\mc{F}(G)$ is a
  generalized eigenfunction to the eigenvalue $\lambda$ if there
  exists a number $n\in\Natural$ such that
\begin{equation}
  \label{eq:GenEigFun}
  (\Delta-\lambda)^n\Phi_\lambda\ =\ 0\ \ .
\end{equation}
  Following \cite{Gotz:2006qp}, let us introduce operators
  $A_n(\lambda)$ which are defined through the relation
\begin{equation}
  A_\lambda^{(n)}\ =\ (\Delta-\lambda)^n-(\Delta_0-\lambda)^n\ \ .
\end{equation}
  In the sequel it will become crucial that each single term of
  $A_\lambda^{(n)}$ contains at least one fermionic
  derivative. After these preparations we consider a
  function $f_\lambda\in\mb{F}(G)$ which is an
  eigenfunction of $\Delta_0$, i.e.\ which satisfies
  $\Delta_0f_\lambda=\lambda f_\lambda$. We then associate
  a family of new functions $\Xi_\lambda^{(n)} f_\lambda$
  to $f_\lambda$ through
\begin{equation}
  \label{eq:Map}
  \Xi_\lambda^{(n)}f_\lambda
  \ =\ \sum_{s=0}^\infty\Bigl[-(\Delta_0-\lambda)^{-n}
  A_\lambda^{(n)}\Bigr]^sf_\lambda\ \equiv \
  \sum_{s=0}^r \left(Q^{(n)}_\lambda\right)^s f_\lambda \ \ .
\end{equation}
  Obviously, the sum truncates after a finite number of terms due
  to the fermionic derivatives which occur in all the operators
  $A_\lambda^{(n)}$. A formal calculation shows furthermore
  that the function $\Xi_\lambda^{(n)}f_\lambda$ is a solution of
  eq.\ \eqref{eq:GenEigFun}. Using the definition \eqref{eq:Map}
  on each of the eigenspaces $\Ker(\Delta_0-\lambda)$ we obtain
  a family of maps $\Xi^{(n)}$ which formally exist
  on the complete function space $\mb{F}(G)$.
\smallskip

  The only problem with the maps $\Xi^{(n)}$ is that they might be
  singular on a certain subspace of $\mb{F}(G)$. In fact, a
  close inspection of our expression \eqref{eq:Map} shows that
  it requires to invert $(\Delta_0-\lambda)$ which may not be
  possible. If this happens, it signals the existence of
  functions in $\mc{F}(G)$ which are not annihilated by
  $(\Delta -\lambda)^n$ for any $\lambda$, and therefore
  implies that some Jordan blocks of the Laplacian must
  have a rank higher than $n$. It may be shown by explicit
  calculation that the family of maps $\Xi^{(n)}$ stabilizes for
  $n>r$ and that the resulting limit map $\Xi$ is well-defined on
  the complete space $\mb{F}(G)$ \cite{Gotz:2006qp}. We then define
  the space $\mc{F}(G)=\text{Im}(\Xi)$ as the image of the auxiliary
  space $\mb{F}(G)$ under $\Xi$. This procedure provides an explicit
  construction of the eigenspaces and Jordan blocks appearing in the
  decomposition \eqref{eq:SpectrumFull}. It should also be recalled
  that the map $\Xi$ acts as an intertwiner between the typical
  subspace of $\mb{F}(G)$ with the reduced action of $\g\oplus\g$
  and the typical subspace of $\mc{F}(G)$ with the full action of
  $\g\oplus\g$ \cite{Gotz:2006qp}. As before, reduced and full
  action refer to the use of the differential operators
  $(\mb{K}^i,\mb{S}_1^a,\mb{S}_{2a},\bar{\mb{K}}^i,\bar{\mb{S}}_1^a,
  \bar{\mb{S}}_{2a})$ and $(K^i,S_1^a,S_{2a},\bar{K}^i,\bar{S}_1^a,
  \bar{S}_{2a})$, respectively.%
\smallskip

  Within the present context we can actually convince ourselves
  that the quadratic Casimir is not diagonalizable on any of the
  projective covers $\mc{P}_\mu$. From the above it is clear that
  every projective cover appears in the decomposition of the
  right regular action on the function space $\mc{F}(G)$ and that
  the corresponding subspace $\mc{M}(\mc{P}_\mu)\otimes\mc{P}_\mu$
  contains functions of the form \eqref{eq:CandFun}. We claim
  that some of the latter must necessarily be proper generalized
  eigenfunctions. In fact, all of them are eigenfunctions
  of $\Delta_0$ with eigenvalue $\lambda = C_\mu/2$. But in
  order for them to be eigenfunctions of $\Delta$, the action
  of $\Xi^{(1)}$ must be well defined. This would require in
  particular that we can invert $\Delta_0 - \lambda$ on
\begin{equation}
  \partial_a\,{R^a}_b (g_B)\,\bartial^b \ \det R(g_B^{-1})
  \,{\bigl[D^{(\mu)}(g_B)\bigr]^\alpha}_\beta\, \theta^1\cdots
  \theta^r\,\bar{\theta}_1\cdots\bar{\theta}_r\ \ .
\end{equation}
  But this is clearly not the case if the Kac module $\mc{K}_\mu$
  contains singular vectors that are reached from the ground
  states through application of a single fermionic generator.
  Hence, we have established our claim for all such labels
  $\mu$. In case the singular vectors of $\mc{K}_\mu$ appear
  only at higher levels, one has to refine the analysis and
  consider also higher order (in the summation index $s$)
  terms in the definition of $\Xi^{(1)}$.

\subsection{Correlation functions}

  By now we have complete control over representation content and
  eigenfunctions of the Laplacian in the weak curvature limit of the
  WZNW model. In addition, we can also compute correlation functions
  in this limit. They are given as integrals over a product of
  functions on the supergroup. Integration is performed with an
  appropriate invariant measure, namely the so-called Haar measure
  $d\mu(g)$ of the supergroup. The easiest way to
  obtain $d\mu$ is to extract it from the invariant metric,
\begin{equation}
  \label{eq:GeoMetric}
  ds^2\ =\ ds_B^2-2\,d\bar{\theta}_a\,{R^a}_b(g_B^{-1})\,d\theta^b\ \ .
\end{equation}
  Here, $ds_B^2$ denotes the standard invariant metric on the bosonic subgroup.
  The total metric has a ``warped'' form since the fermionic bit has an explicit
  functional dependence on the bosonic coordinates $g_B$. We can now obtain the
  desired measure as the superdeterminant of the metric,
\begin{equation}
  \label{eq:Measure}
  d\mu(g)\ =\ d\mu_B(g_B)\,\det\bigl(R(g_B)\bigr)
              \,d\theta^1\cdots d\theta^r\,d\bar{\theta}_1\cdots
               d\bar{\theta}_r
\end{equation}
  where $d\mu_B$ denotes an invariant measure on the bosonic subgroup.
  Once this expression has been written down, we can forget our
  heuristic derivation and check the invariance explicitly. Note that
  the existence of the dilaton \eqref{eq:Dilaton} in the WZNW
  Lagrangian \eqref{eq:FreeFieldInt} is directly related to the presence
  of the factor $\det\bigl(R(g_B)\bigr)$ in the measure.
\smallskip

  Suppose now we are given $N$ generalized eigenfunctions of the 
  Laplacian $\Delta$ on the supergroup. According to the previous
  discussion, the space of eigenfunctions possesses a basis
  of the form
\begin{equation}
  \begin{split}
    \phi^{{\bf a}}_{\mu; {\bf b}}
    &\ =\ \Xi \, f^{{\bf a}}_{\mu;{\bf
    b}} \ = \ \sum_{s=0}^{r} Q^s_\lambda f^{{\bf a}}_{\mu;{\bf b}} \\[2mm]
    \text{where} \ \ f^{{\bf a}}_{\mu;{\bf b}} &\ =\ %
    f^{a_1,\dots,a_s}_{\mu;b_1,\dots,b_t} \ = \
    f_\mu(g_B) \, \theta^{a_1} \cdots \theta^{a_s} \, \bar
    \theta_{b_1} \cdots \theta_{b_t} \ \ .
  \end{split}
\end{equation}
  Here, $f_\mu(g_B)$ are eigenfunctions of the bosonic Laplacian
  $\Delta_0$ with eigenvalue $\lambda$ and
  $\Xi = \Xi^{(r)}, Q_\lambda = Q_\lambda^{(r)}$
  have been defined in eq.\ \eqref{eq:Map}. The $N$-point functions
  of such semi-classical vertex operators are given by the integrals
\begin{equation}
  \begin{split}
    \bigl\langle \phi^{{\bf a_1}}_{\mu_1; {\bf b_1}} \cdots
       \phi^{{\bf a_N}}_{\mu_N; {\bf b_N}} \bigr\rangle &\ =\ %
     \int\!d\mu(g)\ \phi^{{\bf a_1}}_{\mu_1; {\bf b_1}} \cdots
       \phi^{{\bf a_N}}_{\mu_N; {\bf b_N}} \\[2mm]
     &\ =\ %
     \sum_{s_1=0}^r \cdots \sum_{s_N=0}^r \
       \int\!d\mu(g)\ Q_{\lambda_1}^{s_1} f^{{\bf a_1}}_{\mu_1; {\bf b_1}}
       \cdots\,
       Q_{\lambda_N}^{s_N} f^{{\bf a_N}}_{\mu_N; {\bf b_N}}\ \ .
  \end{split}
\end{equation}
  Most of the $(r+1)^N$ terms in this expression vanish due to the
  properties of Grassmann variables and their integration. In fact
  the largest number of non-zero terms that can possibly appear is
  $N\cdot r+1$. This is realized if all eigenfunctions contain terms with
  the maximal number of fermionic coordinates (along with the lower order
  terms that are determined by the action of $Q^s_\lambda$). A
  particularly simple case appears when e.g.\ the first
  eigenfunction $\phi_1 = \phi^{1,2,\dots,r}_{\mu_1;1,2,\dots,r}$
  contains leading terms with $r$ fermions $\theta$ and $\bar
  \theta$ while all others are purely bosonic. In that case,
  the correlator is simply given by
\begin{equation}
  \label{eq:corrMSS}
  \bigl\langle \phi^{1,2,\dots,r}_{\mu_1;1,2,\dots,r} \,
     \phi_{\mu_2} \cdots \phi_{\mu_N} \bigr\rangle \ = \
   \int\! d\mu_B(g_B)\,\det\bigl(R(g_B)\bigr)f_{\mu_1}(g_B)
   f_{\mu_2}(g_B) \cdots f_{\mu_N}(g_B)\ \ .
\end{equation}
  We shall see that very similar results can be established for
  correlators in the full WZNW on type~I supergroups. This is one
  of the subjects we shall address in the next section.

\section{\label{sc:Quantum}The quantum WZNW model}

  After the thorough discussion of its symmetries and its semi-classical
  limit it is now only a small step to come up with a complete solution
  of the full quantum WZNW model. We first show that the free fermion
  resolution gives rise to a natural class of chiral representations.
  Subsequently, we comment on the representation content of the full
  non-chiral theory, sketch the calculation of correlation functions
  and argue that the natural modular invariant partition function can
  be expressed as a diagonal sum over characters
  of Kac modules. We conclude with some speculations about non-trivial
  modular invariants.

\subsection{Chiral representations of the current superalgebra}

  In section \ref{sc:Cov} and \ref{sc:AlgFF} we decribed in some 
  detail the chiral symmetry of WZNW models on supergroups along 
  with their construction in terms of free fermions. Our next aim 
  is to introduce representations $\mc{H}_\mu$ of $\ag$. It is clear 
  that free fermion resolutions provide a natural construction for 
  representations of current superalgebras. What is remarkable, 
  however, is that these representations turn out to be 
  irreducible for generic (typical) choices of $\mu$.  
\smallskip

  According to the results of section \ref{sc:AlgFF} every representation
  of the decoupled system of the bosonic currents $K_B^i$ and the
  fermions $p\theta$ defines a module of the current superalgebra
  via eqs.\ \eqref{eq:FFRes}. In the bosonic part we shall work with
  irreducible representations $\mc{V}_\mu$ of $\ag_0^{\text{ren}}$. If
  the group $G_B$ is compact there will be a finite number of {\em physical}
  representations (the ``integrable'' ones), otherwise one may encounter
  infinitely many of them, including continuous series. We
  identify the physically relevant representations with a subset
  $\Rep(\ag_0^{\text{ren}})\subset\Rep(\g_0)$ within the representation
  labels for the horizontal subalgebra $\g_0$. This is possible since
  the ground states of $\mc{V}_\mu$ form the $\g_0$-module $V_\mu$ upon
  restriction of the $\ag_0^{\text{ren}}$-action to its horizontal subalgebra
  $\g_0$. Note that the curvature of the background geometry leads to
  truncations which imply that $\Rep(\ag_0^{\text{ren}})$ is generally a
  true subset of $\Rep(\g_0)$.\footnote{For $\widehat{su}(2)_k$, for instance,
  the integrable representations are $\lambda=0,1,\ldots,k$ while there
  is no upper bound for unitary $su(2)$-modules.}
  The fermions, on the other hand, admit a unique irreducible
  representation $\mc{V}_F$. The latter is generated from the $SL(2,
  \Complex)$-invariant vacuum $|0\rangle$ by imposing the highest weight
  conditions $(p_a)_n|0\rangle=0$ for $n\geq0$ and $\theta_n^a|0\rangle=0$
  for $n>0$.\footnote{One could include twisted sectors where the moding
  of the fermions is not integer. But then the global supersymmetry would
  not be realized in the WZNW model since there were no zero-modes.}
  The irreducible representations of the product theory therefore
  take the form
\begin{equation}
  \label{eq:Fock}
  \mc{H}_\mu\ =\ \mc{V}_\mu\otimes\mc{V}_F\ \ .
\end{equation}
  Given the free fermion realization \eqref{eq:FFRes}, these spaces admit
  an action of the infinite dimensional current superalgebra $\ag$ as
  defined in \eqref{eq:OPEBB}-\eqref{eq:OPEFF}.
\smallskip

  The generalized Fock modules $\mc{H}_\mu$ provide the proper realization
  of chiral vertex operators as defined around eq.~\eqref{eq:VertexBos}.
  It is indeed evident from our construction that the ground states of
  $\mc{H}_\mu$ transform in the $\g$-module $\mc{K}_\mu$ (recall that the
  ground states of $\mc{V}_\mu$ form the $\g_0$-module $V_\mu$) and that
  they are annihilated by all positive modes of the currents and by the
  zero modes of $S_{2a}(z)$. But there is another and much deeper reason 
  for the relevance of the modules $\mc{H}_\mu$. Observe that the current 
  superalgebra $\ag$ is a true subalgebra of the algebra that is generated 
  from $\ag_0^{\text{ren}}$ and the fermions. Therefore, one might suspect that 
  the spaces $\mc{H}_\mu$ are no longer irreducible with respect to the 
  action of $\ag$. But for generic choices of $\mu$ this is not the case: 
  The action of $\ag$ on $\mc{H}_\mu$ is typically irreducible! This 
  property is in sharp contrast to what happens for standard bosonic
  free field constructions \cite{Feigin1988:MR971497,Gerasimov:1990fi,%
  Bouwknegt:1989jf,Feigin:1990qn,Rasmussen:1998cc} and it characterizes
  the modules $\mc{H}_\mu$ as the natural infinite dimensional lift of
  Kac modules for the finite dimensional Lie superalgebra $\g$. We take
  this observation as a motivation to refer to the generalized Fock
  modules $\mc{H}_\mu$ as Kac modules from now on. Let us emphasize,
  however, that they are constructed in a different manner than those
  of the finite dimensional Lie superalgebra $\g$ in section
  \ref{ssc:KacMod}.
\smallskip

  Since it is a rather crucial issue for the following, we would 
  like to spend some time to establish irreducibility of the 
  representations $\mc{H}_\mu$ for generic labels $\mu$. We shall
  assume for simplicity that the underlying bosonic representation
  $\mc{V}_\mu$ is a highest weight module. The highest weight $\mu$ 
  determines two seemingly different (but in fact equivalent) 
  Verma-like modules of $\ag$. The first of them will be denoted 
  by $\mc{Y}_\mu'$. It is obtained as a product 
  $$ \mc{Y}'_\mu \ = \ \mc{Y}_\mu^{(0,\text{ren})} \otimes 
     \mc{V}_F $$ 
  of the Verma module $\mc{Y}_\mu^{(0,\text{ren})}$ of $\ag_0^{\text{ren}}$
  with the free fermion state space $\mc{V}_F$. We shall consider $\mc{Y}'_\mu$
  as a $\ag$-module. The $\ag$-module $\mc{H}_\mu$ may be recovered from
  $\mc{Y}'_\mu$ by dividing out all the bosonic singular vectors from the 
  $\ag_0^{\text{ren}}$-module $\mc{Y}_\mu^{(0,\text{ren})}$. But there is 
  a second natural Verma-like module $\mc{Y}_\mu$ for $\ag$ which is 
  constructed directly by requiring that all the positive modes as well 
  as the zero-modes $(S_{2a})_0$ annihilate the highest weight, i.e.\ 
  $\mc{Y}_\mu$ is defined without any reference to the free fermion 
  construction of $\ag$. Since the generators $K^i_n, S_{1,n}^a, S_{2b,n}$ 
  and $K^i_{B,n}, \theta^a_n, p_{a,n}$ are in one-to-one correspondence
  with each other, the Verma modules $\mc{Y}_\mu$ and $\mc{Y}'_\mu$ are 
  naturally isomorphic as vector spaces. The natural isomorphism preserves
  the grading  by conformal dimensions. Hence, the characters of
  $\mc{Y}_\mu$ and $\mc{Y}'_\mu$ agree. It is tempting to conjecture that
  $\mc{Y}_\mu$ and $\mc{Y}_\mu'$ are in fact equivalent as $\ag$-modules.
\smallskip

  In order to understand the equality of conformal dimensions we could
  simply refer to the equivalence of energy momentum tensors which
  has been proven in section \ref{sc:AlgFF}. But there is also a more
  pedestrian way of seeing it. In the case of $\ag_0^{\text{ren}}$,
  the current algebra involves the renormalized metric $\kappa-\gamma$
  while the bosonic subalgebra $\ag_0$ of $\ag$ is defined in terms
  of the metric $\kappa$. But according to the Sugawara constructions
  for $\ag_0^{\text{ren}}$ and $\ag$, the respective energy momentum
  tensor requires an additional quantum renormalization of the metric 
  in both cases. This extra renormalization is different as well and
  the final result (the ``fully renormalized metric'') coincides again.
  The previous statement corresponds to the two different ways of
  introducing brackets in the following equation,
\begin{equation}
  \Bigl(\kappa^{ij}-\gamma^{ij}\Bigr)-\frac{1}{2}\,{f^{im}}_n\,{f^{jn}}_m
  \ =\ \kappa^{ij}-\Bigl(\gamma^{ij}+\frac{1}{2}\,{f^{im}}_n\,{f^{jn}}_m\Bigr)\ \ .
\end{equation}
  The first term on both sides refers to the ``classical'' metric and
  the second term describes the quantum renormalization. In addition,
  the effect of the fermions in $\ag$ has to be traded for the
  presence of the dilaton in the $\ag_0^{\text{ren}}$ description.
\smallskip

  Let us now focus on the Verma-like modules $\mc{Y}_\mu$.
  In general, these modules contain singular vectors, certainly
  of bosonic type but possibly also fermionic ones. Our goal
  here is two-fold: First, we would like to argue for a one-to-one
  correspondence of the bosonic singular vectors with those in
  $\mc{Y}_\mu^{(0,\text{ren})}$. Moreover, we would like to show
  that the existence of fermionic singular vectors is an atypical
  event, occurring only for a small subset of weights $\mu$.
\smallskip

  In principle, the structure of singular vectors in the module
  $\mc{Y}_\mu$ can be discussed using a suitable variant of the
  Kac-Kazhdan determinant \cite{Kac1978:MR519631}. For simplicity
  we shall follow a more down-to-earth approach here.
  The existence of a proper submodule $\mc{Y}_\nu$ in the
  representation $\mc{Y}_\mu$ requires that the weight $\nu$ can
  be reached from $\mu$ by (multiple) application of the root
  generators of $\ag$. We may qualify this further with the help
  of two gradings, one with respect to the generator $L_0$\footnote{The
  metric or the level(s), respectively, are assumed to be
  fixed once and for all.} and the other coming from the Cartan
  subalgebra of $\g$ (which is identical to that of $\g_0$). The
  latter implies that the weights $\mu$ and $\nu$ have to be related
  by $\nu=\mu-m\alpha$ where $\alpha$ is a positive root of $\g$
  and $m\in\Integer_{\geq0}$. If the energy direction is considered
  separately, one obtains a necessary condition of the form
\begin{equation}
  \label{eq:PhysKacKazhdan}
  h_{\mu-m\alpha}\ =\ h_{\mu}+nm\ \ ,
\end{equation}
  where $h$ denotes the conformal dimension and the root generator
  belonging to $\alpha$ is assumed to increase the energy by $n$
  units.
\smallskip

  We will investigate condition \eqref{eq:PhysKacKazhdan} for bosonic
  root generators of $\ag$ first. The latter are in one-to-one 
  correspondence with those of $\ag_0^{\text{ren}}$. Since, in addition, 
  the conformal dimensions of highest weight modules $\mc{Y}_\mu$ and
  $\mc{Y}_\mu^{(0,\text{ren})}$ coincide, we conclude that the associated
  decoupling equations \eqref{eq:PhysKacKazhdan} possess the same set 
  of bosonic solutions. We consider this a strong hint that singular 
  vectors in the $\ag_0$-modules $\mc{Y}_\mu^{(0,\text{ren})}\otimes
  \mc{V}_F$ agree with those singular vectors of the $\ag$-modules 
  $\mc{Y}_\mu$ which can be reached by application of bosonic root 
  generators. If we assume this to be true, all bosonic singular 
  vectors are removed when be pass from $\mc{Y}_\mu$ to 
  $\mc{H}_\mu$. Therefore, the singular vectors that remain in 
  $\mc{H}_\mu$ are necessarily fermionic. 
\smallskip

  Let us now look for the existence of potential fermionic singular 
  vectors. We do not intend to formulate any precise rules for when
  they appear, but would like to argue that they must be rare
  compared to their bosonic counterparts. To this end, we recall
  that the conformal dimension $h$ is a quadratic expression of
  the form $h_\mu=\langle\mu,\mu+2\rho\rangle$ (the bracket denoting
  the non-degenerate scalar product that comes with the metric
  \eqref{eq:FulMetric}). Hence, we can always solve eq.\ %
  \eqref{eq:PhysKacKazhdan} for $m$, no matter which bosonic root
  vector $\alpha$ we insert. This ceases to be true for fermionic root
  generators. Since they are nilpotent, eq.\ \eqref{eq:PhysKacKazhdan}
  needs to be solved with $m = 0,1$, something that rarely ever works
  out. Therefore, modules with fermionic singular vectors are called
  atypical. A more systematic study of atypical representations is
  beyond the scope of this article. But the experience with several
  examples suggests that the composition series of the representations
  $\mc{H}_\mu$ is finite and that they possess the same structure as
  the modules of the horizontal subsuperalgebra. In fact, we believe
  that the only possible fermionic
  singular vectors are those that appear on the level of ground
  states and images thereof under the action of certain spectral
  flow automorphisms (see section \ref{sc:SF}).
\smallskip

  Given the structure of the Kac modules \eqref{eq:Fock} it is
  straightforward to derive character formulas and their modular
  properties. Indeed, the characters simply factorize into
\begin{equation} \label{charag}
  \chi_{\mc{H}_\mu}(q)\ =\ \chi_{\mc{V}_\mu}(q)\,\chi_{\mc{V}_F}(q)\ \ .
\end{equation}
  The supercharacter of $\mc{H}_\mu$ has the same product form but
  with the fermionic factor $\chi_{\mc{V}_F}$ being replaced by
  its corresponding supercharacter. Relation \eqref{charag}
  may also be extended to a statement about non-specialized
  characters since the fermions $p_a$ and $\theta^a$ are charged
  under the bosonic generators $K^i$. If $\g_0$ is a simple Lie
  algebra the characters of the unitary $\ag_0^{\text{ren}}$-modules
  $\mc{V}_\mu$ can be looked up in \cite{Kac:1984mq,Kac:1990}. They
  form a finite dimensional unitary 
  representation of the modular group. The character of the fermionic
  representation $\mc{V}_F$, on the other hand, is given by
\begin{equation}
  \chi_{\mc{V}_F}(q)
  \ =\ \Biggl[2q^{\frac{1}{12}}\prod_{n=1}^\infty(1+q^n)^2\Biggr]^r
  \ =\ \Biggl[\frac{\vartheta_2(q)}{\eta(q)}\Biggr]^r\ \ .
\end{equation}
  Under the modular transformation $\tau\mapsto-1/\tau$ the quotient
  $\vartheta_2/\eta$ is simply replaced by $\vartheta_4/\eta$. Hence,
  all the non-trivial information about modular transformations
  resides in the behaviour of the characters for the bosonic
  algebra $\ag_0^{\text{ren}}$. Consequently, the modular properties
  of Kac modules $\mc{H}_\mu$ are under complete control. Even
  though Kac modules do not suffice to build the state space of WZNW
  models on supergroups, the bulk partition function for type~I
  supergroups may be expressed in terms of characters of
  Kac modules (see below). Therefore, modular invariance of
  the bulk partition function is guaranteed as long as it
  involves a summation over the same set of labels as in the
  corresponding bosonic model. The precise construction will be
  explained in more detail in section \ref{sc:SpectrumQM}.
\smallskip

  It remains to work out the characters of atypical irreducible
  representations. The latter are quotients of reducible Kac modules.
  According to our experience with concrete models, the composition series
  of the infinite dimensional Kac modules $\mc{H}_\mu$ of $\ag$ is very
  closely related to that of Kac modules for the horizontal
  subsuperalgebra $\g$. In specific
  examples it is usually straightforward to invert the linear relations
  between characters resulting from
  such a composition series, i.e.\ to express the characters of
  atypical irreducible representations through those of Kac modules.
  A more general approach to this problem using Kazhdan-Lusztig
  polynomials has
  been presented in \cite[Proposition 5.4]{Zou1996:MR1378540} (see
  also \cite{Serganova1998:MR1648107,Brundan2001:MR1937204}). Recently
  it has been shown that the solution for the inversion problem could be
  used to (re)derive the characters of irreducible representations for the
  affine Lie superalgebras $\widehat{sl}(2|1)$ and $\widehat{psl}(2|2)$
  \cite{Gotz:2006qp,Saleur:2006tf}. We expect that this observation
  extends to more general current superalgebras and that it will be
  helpful in the study of modular transformations.
  Representations of affine Lie superalgebras and their behaviour
  under modular transformations have also been studied in
  \cite{Kac:1994kn,Kac:2001,Semikhatov:2003uc}.

\subsection{\label{sc:SF}Spectral flow automorphisms}

  In the previous subsection we have skipped over one rather important
  element in the representation theory of current (super)algebras:
  The spectral flow automorphisms. As we shall recall momentarily,
  spectral flow automorphisms describe symmetry transformations in
  the representation theory of current algebras. Furthermore, they
  seem to be realized as exact symmetries of the WZNW models on
  supergroups, a property that makes them highly relevant for our
  discussion of partition functions below.
\smallskip

  Throughout the following discussion, we shall denote (spectral
  flow) automorphisms of the current superalgebra $\ag$ by $\omega$.
  We shall mostly assume that the action of $\omega$ is consistent
  with the boundary conditions for currents, i.e.\ that it preserves
  the integer moding of the currents. In the
  context of representation theory, any such spectral flow
  automorphism $\omega$ defines a map on the set of (isomorphism
  classes of) representations $\rho:\ag\to\End(V)$ via concatenation,
  $\omega(\rho)=\rho\circ\omega:\ag\to\End(V)$.
\smallskip

  In line with our general strategy, we would like to establish that
  spectral flow automorphisms $\omega$ of the current superalgebra are
  uniquely determined by their action on the bosonic generators.
  A spectral flow automorphism $\omega:\ag_0\to\ag_0$ of the bosonic
  subalgebra $\ag_0$ is, by definition, a linear map\footnote{We refrain
  from introducing a different symbol here such as $\omega_0$.}
\begin{equation}
  \label{cbos}
  \omega\bigl(K^i(z)\bigr)
  \ = \ {(W_0)^i}_j(z)\, K^j(z) + w_0^i\, z^{-1}
\end{equation}
  satisfying certain consistency conditions to be recalled
  below. The map $W_0(z)=z^{\zeta_0}$ is defined in terms of an
  endomorphism $\zeta_0:\g_0\to\g_0$ of the horizontal subalgebra.
  While the eigenvalues of $\zeta_0$ determine how the spectral
  flow shifts the modes of the currents, the vector $w_0^i$ affects
  only the zero-modes. In order to
  preserve the trivial monodromy under rotations around the origin
  we will assume that $W_0(z)$ is a meromorphic function, i.e.\ that
  all the eigenvalues of $\zeta_0$ are integer. Inserting the
  transformation \eqref{cbos} into the operator product expansions
  \eqref{eq:OPEBB} leaves one with the constraints
\begin{equation}
  \label{detzeta0}
  {(\zeta_0)^i}_j
  \ =\ {f^{ik}}_l\,\kappa_{kj}\,w_0^l
\end{equation}
  and
\begin{align}
  \label{W0}
  {(W_0)^i}_k(z)\,{(W_0)^j}_l(z)\,\kappa^{kl}
  &\ =\ \kappa^{ij}&
  &,&
  {f^{ij}}_k\,{(W_0)^k}_l(z)
  &\ =\ {(W_0)^i}_m(z)\,{(W_0)^j}_n(z)\,{f^{mn}}_l\ \ .
\end{align}
  The first equation \eqref{detzeta0} in fact implies that the only
  free parameter is the shift vector $w_0^i$. In the case of a semisimple
  Lie algebra $\g_0$ (which leads to a non-degenerate Killing form)
  this argument can also be reversed and hence it allows to express $w_0^i$
  in terms of $\zeta_0$.
\smallskip

  We would now like to argue that equation \eqref{detzeta0}
  already implies the consistency of the spectral flow (up to the question
  whether $\zeta_0$ has integer eigenvalues), i.e.\ the validity of
  the equations \eqref{W0}. Given the concrete form of $W_0(z)$, it can
  indeed be shown that the two relations \eqref{W0} follow from the
  equations
\begin{align}
  \label{zeta0}
  {(\zeta_0)^i}_k\kappa^{kj}+{(\zeta_0)^j}_l\kappa^{il}
  &\ =\ 0&
  {f^{ij}}_k{(\zeta_0)^k}_l
  &\ =\ {(\zeta_0)^i}_k{f^{kj}}_l+{(\zeta_0)^j}_k{f^{ik}}_l\ \ .
\end{align}
  These relations are in turn just a consequence of \eqref{detzeta0}
  using the invariance of $\kappa^{ij}$ and the Jacobi identity for
  the structure constants.
  Since the same idea will be used again below let us sketch the
  proof of our assertion that the eqs.\ \eqref{zeta0} imply
  the eqs.\ \eqref{W0}. First of all, it is easy to see that one can
  generalize the relations \eqref{zeta0} to powers of $\zeta_0$
  using induction. In the first case, this just yields an alternating
  relative sign, while in the second case it establishes some kind of
  binomial formula. Writing $W_0(z)=\exp(\zeta_0\ln z)$ and expanding
  in powers of $\ln z$ one can then explicitly verify the equations
  for $W_0(z)$. Any vector $w_0^i$ which leads to a matrix $\zeta_0$
  with integer eigenvalues under the identification \eqref{detzeta0}
  will accordingly be referred to as a
  spectral flow automorphism of $\ag$ from now on.
\smallskip

  Given the insights of the previous paragraphs it is now fairly
  straightforward to extend the spectral flow automorphism
  $\omega:\ag_0\to\ag_0$ to the full current superalgebra. To this
  end, we introduce the element
\begin{equation}
  \label{defzeta1}
  \zeta_1\ =\ - R^i\, \kappa_{ij} \, w^j_0 \ \ .
\end{equation}
  It is crucial to observe that this matrix satisfies the
  relation
\begin{equation}
  \label{intzeta1}
  {(\zeta_0)^i}_j\, {(R^j)^a}_c  +  {(\zeta_1)^a}_b\,
  {(R^i)^b}_c \ = \ {(R^i)^a}_b \, {(\zeta_1)^b}_c\ \ ,
\end{equation}
  an analogue of eq.\ \eqref{zeta0}. Following
  the discussion in the bosonic sector, we now introduce a function
  $W_1(z)=z^{\zeta_1}$. Using the same reasoning as in the previous
  paragraph, the equation \eqref{intzeta1} implies
\begin{equation}
  \label{eq:intzeta2}
  {(R^i)^a}_b\,{(W_1)^b}_c(z)
  \ =\ {(W_0)^i}_j(z)\,{(W_1)^a}_b(z)\,{(R^j)^b}_c\ \ .
\end{equation}
  Now we can define the action of the spectral flow automorphism
  $\omega$ on the fermionic currents by
\begin{equation}
\omega\bigl(S_1^a(z)\bigr) \ = \ {(W_1)^a}_b (z)\  S^b_1(z) \ \ \ , \ \
\ \ \omega\bigl(S_{2a}(z)\bigr) \ = \ S_{1b}(z) \,  {({\overline W_1})^b}_a(z)\ \ ,
\end{equation}
  where $\overline W_1$ denotes the inverse of $W_1$. Once more,
  consistency with the operator product expansions of the
  supercurrents is straightforward to verify. The only input
  is the definition \eqref{defzeta1} and the property
  \eqref{eq:intzeta2}.%
\smallskip

  We would also like to argue that the spectral flow symmetry is
  consistent with the free fermion representation \eqref{eq:FFRes}.
  To be more specific, we shall construct an automorphism on the
  chiral algebra of the decoupled system generated by the currents
  $K_B^i(z)$ and the free fermions $p_a(z)$ and $\theta^a(z)$
  that reduces to the expressions above if we plug the transformed
  fields into the defining equations \eqref{eq:FFRes}. In this
  context the most
  important issue is to understand how the renormalization of
  the metric $\kappa \rightarrow \kappa - \gamma$ affects the
  action of the spectral flow. As a consequence of
  eq.\ \eqref{intzeta1} we note that
\begin{equation}
  \label{eq:RenAux}
  {(\zeta_0)^i}_k \, \gamma^{kj} +  {(\zeta_0)^j}_k \, \gamma^{ik}
  \ = \ \tr\bigl( [R^i R^j,\zeta_1]\bigr) \ = \ 0\ \ ,
\end{equation}
  where $\gamma^{ij} = \tr(R^iR^j)$, as before. Consequently,
  the data $\zeta_0$ which gave rise to a spectral flow automorphism
  of $\ag_0$ above, can also be used to define a spectral flow
  automorphism of the renormalized current algebra, i.e.\ of the
  algebra that is generated by $K^j_B$ with operator products given
  in subsection \eqref{sc:AlgFF}. Only the shift vector $w_0^i$ of
  the zero modes needs a small adjustment such that the new spectral
  flow action reads
\begin{equation}
  \label{cBBos}
  \omega\bigl(K^i_B(z)\bigr)
  \ = \ {(W_0)^i}_j(z)\, K^j_B(z) + w_B^i\, z^{-1} \ \ \
  \text{where}  \ \ \ w_B^i \ = \ w^i_0 + \tr\bigl(\zeta_1 R^i\bigr)\ \ .
\end{equation}
  In order to validate that this indeed defines an automorphism
  we need to check the analogue of the condition \eqref{detzeta0}
  for the new metric $\kappa-\gamma$. But this constraint is
  trivially met, using
\begin{equation}
  w_B^i\ =\ (\kappa-\gamma)^{ij}\kappa_{jk}w_0^k\ \ .
\end{equation}
  along with the invariance of both metrics $\kappa$ and
  $\kappa-\gamma$. Note that $\zeta_0$ is not changed and hence
  it has the same (integer) eigenvalues as before.
\smallskip

  In order to obtain an automorphism which is compatible
  with the free field construction we also need to
  introduce the transformations
\begin{align}
  \omega\bigl(p_a(z)\bigr) \ =\ p_b(z) \, {(\overline W_1)^b}_a(z)
  \ \ \ \ , \ \ \ \
  \omega\bigl(\theta^a(z)\bigr) \ = \ {(W_1)^a}_b(z)  \, \theta^b(z)\ \ .
\end{align}
  It is then straightforward but lengthy to check that the
  previous transformations define an automorphism of the algebra
  generated by $p_a$, $\theta^a$ and $K^j_B$ that descends to the
  original spectral flow automorphism $\omega$ of our current
  superalgebra $\ag$. During the calculation one has to be aware
  of normal ordering issues.
\smallskip

  In conclusion we have shown that any spectral flow automorphism
  of the bosonic subalgebra of a current superalgebra (related to
  a Lie superalgebra of type~I) can be extended to the full current
  superalgebra. Furthermore, this extension was seen to be consistent
  with our free fermion resolution. Let us remark that even if we start
  with a spectral flow automorphism $\omega$ preserving periodic
  boundary conditions for bosonic currents, the lifted spectral flow
  $\omega$ does not necessarily have the same property on fermionic
  generators. Only those spectral flow automorphisms $\omega:\ag\to
  \ag$ for which $W_1$ is meromorphic as well seem to arise as symmetries of
  WZNW models on supergroups. Nevertheless, also non-meromorphic
  spectral flows turn out to be of physical relevance. They can be
  used to describe the twisted sectors of orbifold theories,
  see section \ref{sc:ModInv} for details.

\subsection{\label{sc:SpectrumQM}Spectrum and correlation functions}

  Obviously, it is of central importance to determine the partition
  function and higher correlators of WZNW models on supergroups.
  Here we shall explain how the calculation of these quantities
  may be reduced to computations in the corresponding bosonic WZNW
  models. For the torus partition function we will provide a full
  expression in terms of characters of the (renormalized) bosonic
  current algebra.
\smallskip

  All computations in the WZNW model on type I supergroups depart from
  the decoupled theory~\eqref{eq:WZNWDecoupled}. The interaction
  between bosons and fermions is treated perturbatively. What makes
  this approach particularly powerful is the fact that the perturbative
  expansion turns out to truncate after a finite number of terms.
  The order at which the truncation occurs, however, depends on the
  supergroup and the correlator to be computed. As a general rule,
  the number of terms to consider in the perturbative expansion
  increases with the number of vertex operators that are inserted.
\smallskip

  To begin with, let us describe the unperturbed theory~\eqref{eq:WZNWDecoupled}
  with a few concrete formulas. As we proceed it is useful to keep in mind
  that solving the unperturbed theory is a field theoretic analogue of solving
  the truncated Laplace operator $\Delta_0$. Fields in the decoupled theory
  form a space $\mb{H}$ which is a field theoretic version of the semi-classical
  space $\mb{F}(G)$. The state space $\mb{H}$ naturally factorizes into
  bosonic and fermionic contributions,
\begin{equation}\label{eq:Product}
  \mb{H}\ =\ \bigoplus_{\mu\in\Rep(\ag_0^{\text{ren}})}
   \Bigl(\mc{V}_\mu\otimes\mc{V}_F\Bigr)\otimes\Bigl(\mc{\bar{V}}_\mu^\ast
   \otimes\mc{\bar{V}}_F\Bigr)\ \ .
\end{equation}
  For simplicity we assumed that the bosonic part has a charge
  conjugate modular invariant partition function.\footnote{In case
  the consistency of the bosonic theory requires to consider spectral
  flow automorphisms, e.g.\ for non-compact groups, they should also be
  included in the definition of the labels $\mu$.} The fermionic
  representation is unique if we restrict ourselves to the Ramond-Ramond
  sector. In case applications require to include fermionic fields with
  anti-periodic boundary conditions as well, they can be incorporated
  easily. According to eq.\ \eqref{eq:Product}, vertex operators
  of the decoupled theory possess a basis of the form
\begin{equation} \label{Ffield}
   V^{\bf a}_{\mu; \bf b}(z,\bar z) \ \equiv\
     V_{\mu;b_1,\dots,b_t}^{a_1,\dots,a_s} (z,\bar z) \ = \
     V_\mu(z,\bar z)\,  \theta^{a_1}(z) \cdots \theta^{a_s}(z)\,  \bar
     \theta_{b_1}(\bar z) \cdots \bar \theta_{b_t} (\bar z)
\end{equation}
  where $V_\mu$ are vertex operators in the bosonic WZNW model.
  We have noted before that the free fermion theory admits a
  current superalgebra symmetry $\ag\oplus\ag$. The latter is given
  explicitly by the formulas in section \ref{sc:AlgFF}. When analyzed
  with respect to this current superalgebra, the state space $\mb{H}$
  assumes the form
\begin{equation}
  \label{eq:FreeSpace}
  \mb{H}\ =\ \bigoplus_{\mu\in\Rep(\ag)}\mc{H}_\mu\otimes\mc{\bar{H}}_\mu^\ast
\end{equation}
  where $\mc{H}_\mu$ an $\mc{H}^\ast_\mu$ are the Kac modules and their
  duals, as defined in equation~\eqref{eq:Fock}.\footnote{It is the dual
  which is relevant here since we assume the antiholomorphic current
  superalgebra to mimic the differential operators \eqref{eq:DiffR},
  not those in \eqref{eq:DiffL}. Notice that the roles of $S_1^a$ and
  $S_{2a}$ are exchanged in these expressions.} It should be kept in
  mind though that $\mb{H}$ contains an atypical sector (including,
  e.g., $\mc{H}_0\otimes\mc{\bar{H}}_0^\ast$) which is not fully
  reducible. Nevertheless, the zero-modes $L_0$ and $\bar L_0$ of
  the Virasoro-Sugawara fields are fully diagonalizable.
\smallskip

  The true state space $\mc{H}$ of the interacting theory, on the
  other hand, is a field theoretic version of the space $\mc{F}(G)$
  in our minisuperspace theory. In particular, $\mc{H}$ agrees
  with $\mb{H}$ as a graded vector space (with the grading provided
  by the generalized eigenvalues of $L_0$ and $\bar L_0$) and
  even as $\ag_0\oplus\ag_0$-module. But
  when considered as a module of the left and/or right current
  superalgebra, $\mc{H}$ and $\mb{H}$ are fundamentally different.
  While, under the action of e.g.\ the right moving
  currents, $\mb{H}$ decomposes into a sum of typical and atypical
  Kac modules, $\mc{H}$ may be expanded into projectives. The
  corresponding multiplicity spaces, however, do not carry a
  representation of the left moving currents, in contrast to
  what we have seen in eq.~\eqref{eq:FreeSpace}. Instead, atypical
  representations of the left and right moving currents form large
  non-chiral modules $\mc{\hat{I}}_{[\sigma]}$ which entangle
  projective covers in an intricate way.\footnote{Note that the
  structure and number of $\ag$-blocks and hence of the indecomposables
  $\mc{\hat{I}}_{[\sigma]}$ in the field theory may differ from
  that in the minisuperspace theory, see eq.\ \eqref{eq:SpectrumFull}. 
  The relation between the two may be established with the help
  of spectral flow automorphisms.} Now recall that the Virasoro
  element $L_0$ contains the (renormalized) Casimir operator of
  $\g$ as a summand and it agrees with the latter on ground
  states. But since our harmonic analysis revealed that the
  Casimir operator may not be diagonalized in the atypical
  subspace of $\mc{F}(G)$, the same must be true for the action
  of $L_0$ (and $\bar{L}_0$) on $\mc{H}$. This shows that
  supergroup WZNW theories are always logarithmic conformal
  field theories.\footnote{There might exist consistent
  truncations to diagonalizable subsectors for low levels,
  see the discussion in \cite{Saleur:2006tf}. Such phenomena
  appear to be very rare, though.}
\smallskip

  After these remarks, let us address the partition function of
  the theory and its modular invariance. We have stressed above
  that $\mb{H}$ and $\mc{H}$ are isomorphic as
  $\ag_0\oplus\ag_0$-modules. Hence, the partition function of
  the interacting theory agrees with the partition function of
  the decoupled model and both may be written as a sum over
  bilinears of characters of Kac modules.\footnote{Since the
  Cartan subalgebra of $\g$ was assumed to be identical to the
  Cartan subalgebra of $\g_0$ this statement even holds for
  unspecialized characters and partition functions.} Thereby,
  the partition function of WZNW models on type~I supergroups
  takes the form
\begin{equation}
  \label{eq:PartFun}
  Z^G(q,\bar{q})\ =\ Z_{\text{ren}}^{G_B}(q,\bar{q})\cdot
  Z_F(q,\bar{q})\ ,
\end{equation}
  i.e.\ it is obtained as a product of the corresponding partition
  functions of the (renormalized) bosonic model with that of the free
  fermionic system. Each of the two factors corresponds to a well-defined
  and consistent conformal field theory. This shows that our proposal for
  the state space of the supergroup WZNW model yields a suitable partition
  function.
\smallskip

  In theories with fermions one has to distinguish between the purely
  combinatorial partition function which merely counts states and
  the torus vacuum amplitude which is the relevant {\em physical}
  quantity. Since the fermions anti-commute, the latter requires
  an insertion of the fermion number operator $(-1)^{F+\bar{F}}$
  into the trace, thus turning characters into supercharacters.
  In our state spaces, bosonic and fermionic states always come 
  in pairs, causing $Z_F(q,\bar{q})$ to vanish. Actually, this 
  is the usual way in which modular invariance manifests itself 
  in fermionic theories. To avoid dealing with trivial quantities, 
  one may switch to unspecialized characters. The latter lead to 
  a non-vanishing physical partition function. 
\smallskip

  We claim that the expression \eqref{eq:PartFun} is the universal
  partition function for supergroup WZNW models similar to the
  charge conjugate one in ordinary bosonic models. We will indeed
  argue in the following section that this modular invariant can
  be used as the basic building block to derive new, non-trivial
  partition functions using methods that are well-established in
  purely bosonic conformal field theories.
\smallskip

  We wish to conclude this subsection with a few comments on the
  calculation of correlation functions. We have argued above that
  fields in the decoupled and the interacting theory are in
  one-to-one correspondence with each other. In fact, the transition
  from the auxiliary space $\mb{H}$ to the proper state space
  $\mc{H}$ of the supergroup WZNW model is implemented by a
  linear map $\hat{\Xi}:\mb{H}\to\mc{H}$. The latter generalizes
  and extends the map $\Xi$ that we used in the semi-classical
  analysis to identify states in $\mb{F}(G)$ and $\mc{F}(G)$. Let
  us denote the image of the field \eqref{Ffield} under $\hat{\Xi}$
  by $\Phi^{{\bf a}}_{\mu;{\bf b}}$. According to our general
  strategy, correlation
  functions in the interacting theory may be computed through
\begin{equation} \label{eq:exp}
  \bigl\langle \Phi^{{\bf a}_1}_{\mu_1; {\bf b}_1} (z_1,\bar z_1) \cdots
  \Phi^{{\bf a}_N}_{\mu_N; {\bf b}_N} (z_N,\bar z_N)\bigr\rangle \ = \
  \sum_{s=0}^{s_{\text{max}}} \, \frac{1}{s!} \,
  \bigl\langle V^{{\bf a}_1}_{\mu_1; {\bf b}_1} (z_1,\bar z_1) \cdots
  V^{{\bf a}_N}_{\mu_N; {\bf b}_N} (z_N,\bar z_N) \
  \mc{S}^s_{\text{int}}\, \bigr\rangle_0 \ \ ,
\end{equation}
  where the correlators on the right hand side are to be evaluated
  in the decoupled theory. We shall show below that correlators with
  $s \geq s_{\text{max}} = N r$ insertions vanish so that the
  summation over $s$ is finite. Let us also recall that the
  interaction term is given by
\begin{equation}\label{eq:Sint}
  \mc{S}_{\text{int}}\ =\ -\frac{i}{2\pi}\int p_a\,{R^a}_b(g_B)\,
  \bar{p}^b\,dw\wedge d\bar{w}\ \ .
\end{equation}
  Here, the expression ${R^a}_b(g_B)$ should be interpreted as a vertex
  operator of the bosonic WZNW model, transforming in the representation
  $R\otimes R^\ast$.
\smallskip

  There are now two computations to be performed in the decoupled theory.
  First of all, we have to determine correlation functions for the
  bosonic fields $V_{\mu_i}$
  with additional insertions of $s$ vertex operators ${R^a}_b(g_B)$.
  We shall assume the bosonic WZNW model to be solved and hence
  that all these bosonic correlators are known. Let us comment,
  however, that the dependence of such correlation functions on
  the insertion points of ${R^a}_b(g_B)$ is controlled by null
  vector decoupling equations. As usual, these can be exploited
  to derive integral formulas for the required correlation
  functions. We shall not go into any more detail here.

  Instead, let us now comment on the second part of the
  computation that deals with the fermionic sector. Since we are
  dealing with $r$ chiral $bc$ systems at central charge $c=-2$,
  the evaluation is rather standard. According to the usual rules,
  non-vanishing correlators on the sphere must satisfy $\# \theta^a -
  \# p_a = 1$, i.e.\ the number of insertions of a fixed field
  $\theta^a$ must exceed the number of insertions of $p_a$ by one.
  In an $N$-point correlator, any given component $\theta^a$ can
  appear at most $N$ times. The fields $p_a$, on the other hand,
  only emerge from the $s$ insertions of the interaction term.
  Hence, we conclude that all contributions to our correlation
  function with $s \geq N \cdot r$ insertions of
  $\mc{S}_{\text{int}}$ vanish. The non-vanishing terms can
  be evaluated using that
\begin{equation} \label{eq:thpcorr}
 \Bigl\langle\,  \prod_{\nu=1}^{n} p_a(z_\nu) \
           \prod_{\mu=1}^{n+1} \theta^a(x_\mu)
  \,  \Bigr\rangle_0 \ = \
  \frac{\prod_{\nu < \nu'} (z_\nu-z_{\nu'}) \ \prod_{\mu < \mu'}
        (x_\mu-x_{\mu'})}{\prod_\nu \prod_\mu (z_\nu -x_\mu)} \ \
\end{equation}
  and a similar formula applies to $\bar \theta_a$ and $\bar p^a$.
  These expressions can be inserted into the expansion \eqref{eq:exp}.
  Thereby we obtain a formula for the $N$-point
  functions of the WZNW model which presents it as a sum of at
  most $N \cdot r$ terms labeled by an integer $s$. Each
  individual summand involves an integration over $s$ insertion
  points $w_i$. The corresponding integrand factorizes into
  free field correlators of the form \eqref{eq:thpcorr} multiplied
  with a non-trivial $(N+s)$-point function in the bosonic WZNW
  model for the group $G_B$.
\smallskip

  Let us point out that for a given choice of $N$ fields, the
  perturbative evaluation of the correlator may truncate way
  before we reach $s_{\text{max}}$. An extreme example appears
  when all the fields $\Phi_{\mu_i} = \hat{\Xi} V_{\mu_i}, i=2,
  \dots,N,$ are images of purely bosonic fields $V_{\mu_i}$ while
  the first field contains the maximal number of fermionic factors,
  both for left and right movers. In that case, only the term with
  $s=0$ contributes and hence these fields of the WZNW model on the
  supergroup possess the same correlation functions as in the
  bosonic WZNW model, i.e.\ %
\begin{equation}
  \label{eq:corrWZNW}
  \bigl\langle \Phi^{1,2,\dots,r}_{\mu_1;1,2,\dots,r}
    (z_1,\bar z_1) \,
     \Phi_{\mu_2} (z_2,\bar z_2) \cdots \Phi_{\mu_N}(z_N,\bar z_N)
      \bigr\rangle \ = \
  \bigl\langle V_{\mu_1}(z_1,\bar z_1)  \, V_{\mu_2}(z_2,\bar z_2)
  \cdots V_{\mu_N}(z_N,\bar z_N) \bigr\rangle_0
\end{equation}
  where the correlation function on the right hand side is to be
  evaluated in the bosonic WZNW model. The result is a direct
  analogue of the corresponding formula \eqref{eq:corrMSS} in
  the minisuperspace theory.

\subsection{\label{sc:ModInv}Some comments on non-trivial modular invariants}

  During the course of the previous sections we frequently assumed
  that the bosonic subgroup $G_B\subset G$ was compact and
  simply-connected. On a technical level, this condition is required
  in order to render the matrix $R(g_B)$ well-defined which entered
  the expression for the differential operators implementing the
  isometries of $G$ on the function space $\mc{F}(G)$. On the
  other hand this choice automatically limited our considerations to
  WZNW models with (the analogue of a) charge conjugate modular
  invariant. In this subsection we would like to sketch how
  such a restriction may be overcome.
\smallskip

  Let us recall the situation for bosonic WZNW models first.
  It is well-known that a non-simply-connected group manifold $G_0$
  can be described geometrically as an orbifold $\tilde{G}_0/\Gamma$
  where $\tilde{G}_0$ is the universal covering group and
  $\Gamma\cong\pi_1(G_0)\subset\mc{Z}(\tilde{G}_0)$ is a subgroup of
  its center. The simplest example is $SO(3)=SU(2)/\Integer_2$.
  In conformal field theory, orbifolds of the previous type are
  implemented by means of a simple current extension of the theory
  with charge conjugate modular invariant \cite{Kreuzer:1994tf}
  (see also \cite{Fuchs:2004dz}).
  This construction of the $G_0$ WZNW model rests on the fact
  that the $\tilde G_0$ model contains sufficiently many
  simple currents, one for each element in the center $\mc{Z}
  (\tilde{G}_0)$. Incidently, these are in one-to-one
  correspondence with (spectral flow) automorphisms of the
  current algebra $\ag_0$. Such simple current extensions
  exhaust all modular invariants related to the current
  algebra $\ag_0$, apart from some exceptional cases at
  low levels.
\smallskip

  Now it has been shown in \cite{Rogers:1980mu} that the global
  topology of a Lie {\em super}group is completely inherited from
  that of its bosonic subgroup. Consequently, given a supergroup
  $G$ with bosonic subgroup $G_0=\tilde{G}_0/\Gamma$, there exists
  a covering supergroup $\tilde{G}$ with bosonic subgroup $\tilde{G}_0$,
  and one has $G=\tilde{G}/\Gamma$. Note that central elements in
  $\tilde{G}_0$ are also central in $\tilde{G}$. Having
  constructed the WZNW model on the covering supergroup $\tilde
  G$, we would like to divide by $\Gamma$. But, as we have just
  stated, elements of $\Gamma$ can all be identified with elements
  in the center of the bosonic subgroup $\tilde G_0$. Therefore,
  they label certain simple currents of the $\tilde G_0$ WZNW
  model. As indicated in the previous paragraph, we may think
  of these simple currents as (equivalence classes of) spectral
  flow automorphisms of $\ag_0$. According to the results of
  subsection \ref{sc:SF}, all such spectral flow automorphisms
  may be extended from $\ag_0$ to the current Lie superalgebra
  $\ag$, in a way that is even consistent with the free fermion
  construction. Consequently, the elements of our designated
  orbifold group $\Gamma$ label a certain set of spectral flow
  automorphisms of $\ag$. It is the action of {\em these} spectral
  flow automorphisms that one has to use in order to construct the
  orbifold CFT belonging to the supergroup $G=\tilde{G}/\Gamma$.
\smallskip

  Our discussion so far has been fairly abstract and we would like
  to flesh it out a bit more. Actually, the details of the orbifold
  construction are not much different from what is done in bosonic
  models. For simplicity, let us assume that $\Gamma$ is cyclic and
  of finite order. We shall denote the generating element by $\gamma$.
  In order to illustrate the relation between orbifolds and spectral
  flow automorphisms, we depart from the conventional orbifold
  approach. Namely, we include (chiral) twisted sectors on which
  the supercurrents $X$ satisfy boundary conditions of the form
  \begin{equation}
  X(e^{2\pi i}z)\ =\ \bigl(\gamma(X)\bigr)(z)\ \ .
  \end{equation}
  There exists a basis $X_\sigma$, on which $\gamma$ acts
  diagonally as a multiplication with some phase $\exp(2\pi i
  \gamma_\sigma)$. If $\gamma_\sigma$ is an integer, then $X_\sigma$
  has integer moding in the twisted sector, otherwise its modes
  are rational. All these twisted sectors emerge by acting with
  certain (meromorphic or not) spectral flow automorphisms on the
  untwisted representations (see subsection \ref{sc:SF}). The
  discussion of the previous paragraph supplied us with the
  relevant set of spectral flow automorphisms and hence with
  a list of chiral sectors to be incorporated in the
  construction of the $G = \tilde G/\Gamma$ orbifold theory.
  Sectors of the full non-chiral theory are obtained by independent
  action of spectral flow automorphisms on left and right-movers in
  the parent theory on $\tilde G$. Therefore, even meromorphic
  spectral flows lead to new non-chiral sectors, though these
  are put together from untwisted representations of the left
  and right movers. All this has been worked out for many
  interesting bosonic models, such as e.g.\ the $SO(3)=SU(2)/
  \Integer_2$ WZNW model. WZNW models on non-simply-connected
  supergroups are no harder to deal with.\footnote{
  The $SO(3)$ theory also shows that the orbifold construction
  might suffer from obstructions, depending on the choice of
  the level. A more detailed treatment of such issues for
  supergroup orbifolds is left for future work.}
\smallskip

  Let us finally comment on the connection of the algebraic
  orbifolds with the Lagrangian picture. Looking at our free
  field resolution \eqref{eq:FreeFieldInt} one might have had
  the naive idea to replace the bosonic model by its orbifold
  and then to add fermions and interaction terms in the same
  way as before. But this is not at all what we suggest to do.
  In particular, the orbifold group $\Gamma$ need not be a
  symmetry of the interaction term if there is no action on
  the fermions. Even worse, the vertex
  operator ${R^a}_b(g_B)$ occuring in the interaction may not
  be part of the spectrum of the purely bosonic WZNW model. As
  a consequence, the perturbed correlation functions with
  insertions of this operator are not well-defined. This
  happens, for example, if we try to supersymmetrize the
  bosonic group $SO(3)\times U(1)$. The fermions of the
  extended model with $su(2|1)$ symmetry transform in the
  spin $1/2$ representation of $SU(2)$ which does not descend
  to a representation of $SO(3)=SU(2)/\Integer_2$. Hence, it
  is absolutely crucial to depart from the full $SU(2|1)$
  WZNW model and to divide the full orbifold action on
  both bosonic and fermionic variables.

\section{\label{sc:triplet}Lessons for other logarithmic CFTs}

  Various logarithmic conformal field theories have been considered
  in the literature. The best studied examples are the triplet models
  in which the conformal symmetry is extended by a triplet of currents,
  each having spin $h=2p-1$ \cite{Kausch:1990vg}. For most of these
  algebras only chiral aspects have been investigated so far, but in 
  case of $p=2$, Gaberdiel and Kausch have been able to come up with a
  consistent local theory \cite{Gaberdiel:1998ps}. The extended
  chiral symmetry of the triplet models is denoted by $\mc{W}_{1,p}$.
  The latter are believed to be part of a family of
  more general $\mc{W}$-algebras $\mc{W}_{q,p}$ where $p$ and $q$ are
  co-prime. All of these possess interesting indecomposable
  representations. Their representation theory is particularly
  well understood for $q=1$, see \cite{Fuchs:2003yu}
  and references therein.
\smallskip

  This final section has two aims. First of all we would like to
  illustrate that the existing results on the representation theory
  of $\mc{W}_{1,p}$-algebras and the local triplet model (for $p=2$)
  fit very nicely into one common picture with the logarithmic WZNW
  models on type I supergroups. But given the remarkable progress with
  the latter, and in particular with the construction of infinitely
  many families of new local non-chiral models, our results lead
  to a number of interesting predictions on $\mc{W}_{q,p}$-algebras
  and the associated local logarithmic conformal field theories.

\subsection{Chiral representation theory}

  Let us begin this subsection by reviewing some results on the
  representation theory of $\mc{W}_{1,p} = \mc{W}(p)$ (see
  \cite{Fuchs:2003yu} and references therein). This chiral algebra
  is known to admit $2p$ irreducible highest weight representations
  $\mc{V}_s^\pm$ where $s=1,\ldots,p$. While $\mc{V}^\pm_p$ do not
  admit non-split extensions, all other $2(p-1)$ representations
  appear in the head of the following indecomposables,\footnote{These
  diagrams have to be read as follows: To the right we write the
  maximal fully reducible submodule. Everything left of the
  rightmost arrow describes the quotient module of the original
  module with respect to the submodule mentioned before. One can
  then proceed iteratively to define the whole diagram.}
\begin{equation}
  \mc{R}^\pm_s: \ \quad\mc{V}^\pm_{s}\ \to \ 2\mc{V}^\mp_{p-s}
   \ \to\ \mc{V}^\pm_s
\end{equation}
  where $s$ runs from $s=1$ to $s=p-1$. Hence the representations
  $\mc{V}^\pm_p$ can be considered typical whereas all others are
  atypical.\footnote{We use the qualifiers ``atypical'' and
  ``typical'' only to clarify the analogy to the supergroup WZNW
  models. In contrast to the latter, the atypical representations
  are obviously the generic ones for the algebra $\mc{W}(p)$.}
  Moreover, the indecomposables $\mc{R}^\pm_s$ are the projective
  covers of the atypicals $\mc{V}^\pm_s$ and play the role of
  the representations $\mc{P}$ in section \ref{sc:ProjMod}.
  The typical modules $\mc{V}^\pm_p$ are projective as well, in
  agreement with results on the fusion for $\mc{W}(2)$
  representations, see \cite{Gaberdiel:1996np}. The fusion
  rules of $\mc{W}_{p,q}$-models have recently been addressed 
  in \cite{Flohr:2007jy}.
\smallskip

  The representation theory of $\mc{W}(p)$-algebras also contains
  analogues of our Kac modules for atypical representations. These
  have the form
\begin{align}
  \mc{K}^\pm_s: &\quad\mc{V}^\pm_s\ \to\ \mc{V}^\mp_{p-s}
\end{align}
  where $s=1,\dots,p-1$. In view of the role they are going to play
  we will simply refer to the representations $\mc{K}_s^\pm$ as
  ``Kac modules'' as well.\footnote{Using the analogy to the
  Kazhdan-Lusztig dual quantum group, they have been called Verma
  modules in \cite{Feigin:2005zx}.} They are obtained as quotients of the
  projective covers $\mc{R}^\pm_s$. For the typical representations
  $\mc{V}^\pm_p$, the associated irreducibles, ``Kac modules'' and
  projective covers all coincide. In this sense, we shall also write
  $\mc{K}^\pm_{p}=\mc{V}^\pm_p=\mc{R}_p^\pm$, just
  as for typical representations of Lie superalgebras.
  Furthermore, among the quotients of the projective covers one
  can also find $4(p-1)$ ``zig-zag'' modules, containing three
  irreducible representations each. It seems likely, that these
  are just the first few examples among an infinite series of
  zig-zag representations of $\mc{W}(p)$, in close analogy to
  representations of the Lie superalgebra $gl(1|1)$ (see e.g.\
  \cite{Gotz:2005jz}). The main difference between $gl(1|1)$ and
  $\mc{W}(p)$ zig-zag modules is that the constituents of the
  former are pairwise inequivalent. Zig-zag modules of $\mc{W}(p)$,
  on the other hand, are built from a pair of irreducibles, each
  appearing with some multiplicity. This opens the possibility to
  close zig-zag modules of $\mc{W}(p)$ into rings. Representations
  of all these different shapes were found and investigated for
  the quantum groups \cite{Erdmann:2004,Feigin:2005xs} which are
  dual to $\mc{W}(p)$, in the sense of Kazhdan-Lusztig duality.
\smallskip

  Let us also compare some further properties of $\mc{W}(p)$-modules
  with those we discussed for Lie superalgebras of type~I. For
  example, we have pointed out that all projective modules of
  type~I superalgebras possess a ``Kac composition series''. The
  same is true for the projective covers $\mc{R}^\pm_s$,
\begin{equation}
  \mc{R}^\pm_s:\ \quad\mc{K}^\pm_s\ \to\ \mc{K}^\mp_{p-s}
  \qquad(\text{for }s<p)\quad,
  \qquad\qquad
  \mc{R}_p^\pm=\mc{K}_p^\pm\ \ .
\end{equation}
  Moreover, we also observe that the multiplicities
  in the ``Kac composition series'' of indecomposable projective
  covers (reducible and irreducible) and those of irreducible
  representations in the composition series of
  ``Kac modules'' are related by
\begin{equation}
  (\mc{R}_\mu:\mc{K}_\nu)\ =\ [\mc{K}_\nu:\mc{V}_\mu]\ \ .
\end{equation}
  This establishes an analogue of the reciprocity theorem
  \eqref{eq:BGG} that has been an important ingredient in
  our description of supergroup WZNW models and, in particular,
  in exhibiting its modular invariance.
\smallskip

  The agreement between algebraic structures in the representation
  theory of Lie superalgebras and of symmetries in minimal
  logarithmic conformal field theories is remarkable.
  But let us think
  ahead and see what Lie superalgebras may teach us for future
  studies of indecomposable $\mc{W}$-algebra representations.
  While irreducible representations and their projective covers
  are certainly central objects for all Lie superalgebras, some
  of their properties may differ considerably from what we have
  seen in the case of type~I. We have pointed out already that
  the existence of a ``Kac composition series'' (or a similar
  {\em flag}) for projectives and the reciprocity property
  \eqref{eq:BGG} do not hold for more general Lie superalgebras.
  Hence, these features of $\mc{W}(p)$-modules
  should not be expected to carry over to more
  general $\mc{W}$-algebras either. In fact, numerical results
  of \cite{Eberle:2006zn} may indicate that violations even occur
  for $\mc{W}_{q,p}$ with $p,q\neq 1$. Furthermore, the tensor
  products for irreducible representations of Lie superalgebras
  can develop a remarkable complexity. In this sense, the Lie
  superalgebra $gl(1|1)$ is rather well-behaved. Representations
  of $psl(2|2)$, for example, are much less tame. In particular,
  tensor powers of its adjoint representation lead to an
  infinite series of indecomposables (see \cite{Gotz:2005jz} for
  details). The similarities between representations of
  $gl(1|1)$ and $\mc{W}_{1,p}$ suggest that the latter may
  also be rather unusual creatures in the zoo of
  $\mc{W}$-algebras. In fact, when it comes to the features of
  fusion, the algebras $\mc{W}_{q,p}$ may have much more generic
  properties, resembling very closely those of
  $psl(2|2)$.\footnote{A certain similarity between the
  representation theory of $\mc{W}_{q,p}$ (or rather its dual
  quantum group) and $psl(2|2)$ is suggested by the structure
  of their respective projective covers, cf.\ e.g.\ Figure 7 of
  \cite{Feigin:2006iv} with eq.\ (2.12) of \cite
  {Gotz:2005ka}.}

\subsection{Local logarithmic conformal field theories}

  Regarding the construction of local field theories, the
  progress with WZNW models on supergroups has been significantly
  faster than for minimal logarithmic CFTs. In fact, only the minimal
  triplet model associated with $\mc{W}(2)$ has been constructed in
  all detail \cite{Gaberdiel:1998ps}. Imposing locality constraints
  on correlation functions, the state space
  $\mc{H}$ of this model was shown to have the form
\begin{equation}
  \label{triplet}
  \mc{H}
  \ =\ \mc{I}_1\oplus\bigl(\mc{V}_2^+\otimes\mc{\bar{V}}_2^+\bigr)
  \oplus\bigl(\mc{V}_2^-\otimes\mc{\bar{V}}_2^-\bigr)\ \ .
\end{equation}
  Here, $\mc{V}^\pm_2$ are the typical modules of $\mc{W}(2)$,
  in view of their conformal dimensions previously also denoted by
  $\mc{V}_{-1/8}$ and $\mc{V}_{3/8}$, and $\mc{I}_1$ is a complicated
  non-chiral indecomposable (denoted by $\mc{R}$ in
  \cite{Gaberdiel:1998ps}) which was obtained originally as a certain
  quotient of the space
  $\bigl(\mc{R}^+_1\otimes\bar{\mc{R}}^+_1\bigr)\oplus\bigl(\mc{R}^-_1
  \otimes\bar{\mc{R}}^-_1\bigr)$.
  The module $\mc{I}_1$ is known to possess the following composition series
\begin{equation}
  \mc{I}_1:\quad\bigl(\mc{V}_1^+\otimes\bar{\mc{V}}_1^+\bigr)\oplus
  \bigl(\mc{V}_1^-\otimes\bar{\mc{V}}_1^-\bigr)
  \to2\bigl(\mc{V}_1^+\otimes\bar{\mc{V}}_1^-\bigr)\oplus2
  \bigl(\mc{V}_1^-\otimes\bar{\mc{V}}_1^+\bigr)
  \to\bigl(\mc{V}_1^+\otimes\bar{\mc{V}}_1^+\bigr)\oplus
  \bigl(\mc{V}_1^-\otimes\mc{V}_1^-\bigr)\ \ ,
\end{equation}
  where we used the correspondence $\mc{V}^+_1=\mc{V}_0$ and
  $\mc{V}^-_1=\mc{V}_1$ for the atypical irreducibles of $\mc{W}(2)$.
  When acting with elements of either the left or right chiral algebra
  only, $\mc{H}$ decomposes into a sum of projectives, each appearing
  with infinite multiplicity. The individual multiplicity spaces cannot
  be promoted to representation spaces of the commuting chiral algebra,
  but they come equipped with a grading that is given by the (generalized)
  eigenvalues of $L_0$ or $\bar L_0$. When considered as graded vector
  spaces, they coincide with the graded carrier spaces of irreducible
  representations. All this is very reminiscent of what we found in
  eq.\ \eqref{eq:SpectrumLR} while studying the harmonic analysis on
  supergroups.%
\smallskip

  Carrying on with the comparison between the triplet model and
  WZNW models on supergroups, we also observe that the
  composition series of the state space \eqref{triplet} agrees
  with that of the module $\bigl(\mc{K}_1^+\otimes\bar{\mc{K}}_1^+\bigr)
  \oplus\bigl(\mc{K}_1^-\otimes\bar{\mc{K}}_1^-\bigr)$. Hence, the partition
  function of the triplet model can be expressed as
\begin{equation}
  \label{eq:LogPF}
  Z(q,\bar{q})
  \ =\ \sum_{i=1,2}\sum_{\eta=\pm}\chi_{\mc{K}_i^\eta}(q)\,
  \bar{\chi}_{\mc{K}_i^\eta}(\bar{q})\ \ .
\end{equation}
  This result is reminiscent of what we found for supergroup
  WZNW models in section \ref{sc:SpectrumQM}. Note that the
  modular transformation behaviour for characters of Kac modules
  is rather simple which makes it easy to check that $Z(q,\bar q)$
  is modular invariant. In comparison, the transformation behaviour
  of characters belonging to atypical irreducible representations
  of $\mc{W}(2)$ is rather involved \cite{Gaberdiel:1996np}, just
  as for current superalgebras.
\smallskip

  The striking similarities between the local triplet theory and
  the harmonic analysis on supergroups suggest some far reaching
  generalizations, in particular concerning the state space of
  a wide class of local logarithmic conformal field theories.
  Let us denote the irreducible representations of some chiral
  algebra $\mc{W}$ by $\mc{V}_a$ and their projective covers by
  $\mc{P}_a$. For typical representations the latter agree (by
  definition) with
  the irreducibles. We also introduce the symbol $V_a$ when
  $\mc{V}_a$ is considered merely as an $L_0$-graded vector
  space. Given this notation, we propose that a local logarithmic
  conformal field theory with symmetry $\mc{W}$ can be constructed
  on the state space
\begin{equation}
  \mc{H} \ = \ \bigoplus_a \ V_a  \otimes \bar{\mc{P}}_a\ \ .
\end{equation}
  Our proposal describes the state space of the conjectured local
  theory as a graded representation space for $\mc{W}$.
  The extension to the full $\mc{W} \otimes \bar{\mc{W}}$ is
  severely constrained by requiring symmetry with respect to
  an exchange of left and right chiral algebras. Concerning
  the implications for $\mc{W}(p)$-models it is interesting
  to observe that the same structures were found in the
  regular representation of the dual quantum group, see
  \cite{Feigin:2005zx}, page 24,  and compare with eq.\ %
  (2.9) in \cite{Saleur:2006tf}. Let us point
  out that local theories may probably also be built on other state
  spaces. Examples are given by the orbifold models we described in
  section \ref{sc:ModInv} or by some exceptional truncations of WZNW
  models on simply connected Lie supergroups (see \cite{Saleur:2006tf}
  for a few examples).
\smallskip

\begin{figure}
\begin{equation*}
{\xymatrixrowsep{12pt}
\xymatrixcolsep{12pt}
  \xymatrix{ & \mc{V}_s^+\otimes\mc{V}_s^+\ar[dr]\ar[drrr]\ar[drrrrr]\ar[dl]& & & & \mc{V}_{p-s}^-\otimes\mc{V}_{p-s}^-\ar[dr]\ar[dl]\ar[dlll]\ar[dlllll] \\
      \mc{V}_{p-s}^-\otimes\mc{V}_s^+ \ar[dr]\ar[drrrrr] & & \mc{V}_s^+\otimes\mc{V}_{p-s}^- \ar[dl] \ar[drrr] & & \mc{V}_s^+\otimes\mc{V}_{p-s}^- \ar[dlll] \ar[dr] & & \mc{V}_{p-s}^-\otimes\mc{V}_s^+ \ar[dl]\ar[dlllll] \\
             & \mc{V}_{p-s}^-\otimes\mc{V}_{p-s}^- & & & & \mc{V}_s^+\otimes\mc{V}_s^+}}
\end{equation*}
  \caption{\label{fig:I}The structure of the non-chiral representation $\mc{I}_s$.}
\end{figure}
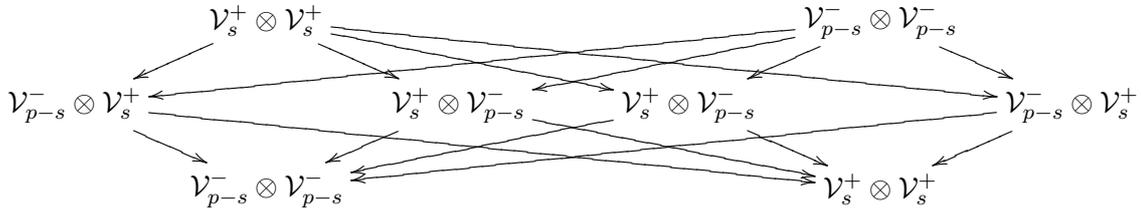

  Before we conclude we would like to go one step beyond the previous
  analogy and to propose a more detailed conjecture for the natural
  state space of the $\mc{W}(p)$ triplet models for arbitrary $p$.
  In a straightforward extension of the result \eqref{triplet} for
  $p=2$ we believe that a local theory may be built on the space
\begin{equation}
  \mc{H}\ =\ \bigoplus_{s\neq p}\mc{I}_s\oplus\bigoplus_{\eta=\pm}
  \mc{V}_p^\eta\otimes\bar{\mc{V}}_p^\eta\ \ .
\end{equation}
  The non-chiral indecomposable representations occurring here
  have the composition series
\begin{equation*}
  \mc{I}_s:\quad\bigl(\mc{V}_s^+\otimes\bar{\mc{V}}_s^+\bigr)
\oplus\bigl(\mc{V}_{p-s}^-\otimes\bar{\mc{V}}_{p-s}^-\bigr)
  \to2\bigl(\mc{V}_s^+\otimes\bar{\mc{V}}_{p-s}^-\bigr)\oplus2
  \bigl(\mc{V}_{p-s}^-\otimes\bar{\mc{V}}_s^+\bigr)
  \to\bigl(\mc{V}_{p-s}^-\otimes\bar{\mc{V}}_{p-s}^-\bigr)
  \oplus\bigl(\mc{V}_s^+\otimes\mc{V}_s^+\bigr)
\end{equation*}
  which coincides with the composition series of 
  $\bigl(\mc{K}_s^+\otimes\mc{K}_s^+\bigr)
  \oplus\bigl(\mc{K}_{p-s}^-\otimes\mc{K}_{p-s}^-\bigr)$. 
  Consequently, our proposal is manifestly modular invariant
  since the partition function can be written as a sum over all
  ``Kac modules'', just as in eq.\ \eqref{eq:LogPF}. Figure \ref{fig:I}
  provides an alternative 2-dimensional picture of the indecomposables 
  $\mc{I}_s$. In this form the similarities with analogous pictures for 
  $gl(1|1)$ and $sl(2|1)$ \cite{Schomerus:2005bf,Saleur:2006tf} and for the 
  quantum group dual of $\mc{W}(p)$-models \cite{Feigin:2005zx} are clearly 
  displayed. 
\smallskip

  The particular relevance of projective modules for local bulk
  theories is one of the main outcomes from the study of WZNW
  models on supergroups, see also \cite{Schomerus:2005bf,Gotz:2006qp,%
  Saleur:2006tf}. Their role for logarithmic extensions of minimal
  models was also emphasized in \cite{Feigin:2006iv,Feigin:2006xa},
  mostly based on studies of the dual quantum group. It seems
  worth pointing out, though, that for quotients of supergroups,
  projective modules might not play such a prominent role, even
  though some of them are likely to be logarithmic as well.
  Similarly, boundary spectra in logarithmic conformal field
  theories are known to involve atypical irreducibles as well
  as projectives. For the triplet model, boundary conditions
  with an atypical irreducible spectrum of boundary operators
  were exhibited in the recent work of Gaberdiel and Runkel
  \cite{Gaberdiel:2006pp}. Studies of branes on supergroups
  confirm the existence of such boundary spectra and they
  provide a beautiful geometric explanation \cite{Creutzig:2007}.

\section{Outlook and open questions}

  In our paper we presented the main ingredients for a complete solution
  of arbitrary supergroup WZNW models based on basic Lie superalgebras
  of type~I. All our results relied on a free fermion resolution of the
  underlying current superalgebra which allowed to keep the bosonic
  subsymmetry manifest in all expressions we encountered, i.e.\ in
  action functionals, representations, correlation functions and other
  quantities. On the level of the Lagrangian we showed that
  the original WZNW Lagrangian could be written as a sum of a WZNW model
  for the bosonic subgroup with renormalized metric and possibly a
  dilaton, the action for a set of free fermions and an interaction term
  which couples the fermions to a vertex operator of the bosonic model.
  The usefulness of this construction has also been demonstrated in the
  full quantum theory, e.g.\ when we reconstructed the current
  superalgebra in terms of the corresponding bosonic current algebra and
  free fermions.
\smallskip

  In order to solve the WZNW model we first focused on its semi-classical, or
  small curvature limit which allowed to reduce the construction of the
  space of ground states to a problem in harmonic analysis on a supergroup.
  We could confirm previous observations
  \cite{Huffmann:1994ah,Schomerus:2005bf,Gotz:2006qp,Saleur:2006tf} that the
  space of functions splits into two qualitative very different sectors.
  First of all, there exists a typical sector which
  decomposes into a tensor product of irreducible typical representations
  under the action of the supergroup isometry $\g\oplus\g$. On this subspace,
  the Laplacian is
  fully diagonalizable and its eigenvalues are determined by a specific
  quadratic Casimir of $\g$. In addition, the space of functions on a
  supergroup always exhibits an atypical sector consisting of projective
  covers entangled in a complicated way such that the resulting non-chiral
  modules cannot be written as (a direct sum of) tensor product representations.
  In this sector the Laplacian is not diagonalizable and the necessity for
  a non-trivial entanglement may eventually be traced back to the fact that
  the left and right regular action lead to the same expression for the
  Laplacian. We wish to emphasize that our derivation of the spectrum
  has been very general and just relied on the validity of a reciprocity
  theorem proven by Zou and Brundan \cite{Zou1996:MR1378540,Brundan2004:MR2100468}.
\smallskip

  Starting from this semi-classical truncation it has been argued that
  all its interesting features persist in the full quantum theory. In
  particular, the full state space of the WZNW model is still composed
  of a typical and an atypical sector. Again, the representations in
  the latter do not factorize and the dilatation operators
  $L_0$ and $\bar{L}_0$ may not be diagonalized. Since the vacuum
  representation is always atypical this automatically implies the
  existence of a logarithmic partner of the identity field and makes
  supergroup WZNW models genuine examples of logarithmic
  conformal field theories.
\smallskip

  It should be noted that, in comparison to ordinary free
  field constructions \cite{Feigin1988:MR971497,Gerasimov:1990fi,%
  Bouwknegt:1989jf,Feigin:1990qn,Rasmussen:1998cc} which are
  based on a choice of an abelian subalgebra, our free
  fermion resolution is much easier to deal with. In particular,
  the representations of the current superalgebra obtained from the
  generalized Fock spaces \eqref{eq:Fock} are typically irreducible.
  This observation lets us suggest that these representations are the
  proper generalization of Kac modules in the infinite dimensional
  setting. Furthermore, there was no need of introducing various
  screening charges and BRST operators, a simplifying feature that
  reflects itself in the calculation of correlation functions. The
  latter could be reduced to a perturbative but finite expansion in
  terms of correlation functions in the product of a purely bosonic
  WZNW model with renormalized metric and a theory of free fermions.
\smallskip

  Finally, we commented on possible partition functions and we explained
  how they are constructed as a product of partition functions for the
  constituents in our free fermion resolution. This rather simple behavior
  is rooted in the fact that traces are insensitive to the composition
  structure of representations. Hence, the full WZNW theory possesses
  the same partition function as the decoupled free fermion
  theory in which products of (reducible) Kac modules appear
  instead of projective covers. Taking this assertion for granted,
  the torus modular invariance of our theory is satisfied automatically.
  It might be helpful to add that torus partition functions of many
  non-rational bosonic conformal field theories, e.g.\ of Liouville theory
  or of the $H^3_+$ model, are equally insensitive to the interaction. This
  does certainly not imply that the theories are trivial, neither in case
  of non-rational conformal field theories, nor for WZNW models on supergroups.
\smallskip

  In the last section of this work we placed our new results on
  chiral and non-chiral aspects of supergroup WZNW models in the
  context of previous and ongoing work on other logarithmic
  conformal field theories, in particular on logarithmic extensions
  of minimal models. The similarities are remarkable and provide some
  novel insight that helps to separate generic properties of
  logarithmic conformal field theories from rather singular
  coincidences. As an application of the analogies we
  conjectured a precise formula for the state space of a fully
  consistent local theory based on an arbitrary
  chiral algebra. It adopts a particularly nice shape
  for the minimal logarithmic $\mc{W}(p)$-theories.%
\smallskip

  Working with supergroup WZNW models has two important advantages
  over the consideration of non-geometric logarithmic conformal
  field theories. Concerning the study of
  chiral aspects, the close link between the current superalgebra
  $\ag$ and its horizontal subsuperalgebra $\g$ provides a rather
  strong handle on the representation theory of $\mc{W} =
  \ag$. In fact, since the representation theory of $\g$ is
  under good control, the same is true for its affine extension
  $\ag$. Even though we have not really pushed this to the level
  of mathematical theorems, there is no doubt that rigorous
  results can be established along the lines of our discussion.
  For some particular examples, this has been carried out
  already \cite{Schomerus:2005bf,Gotz:2006qp,Saleur:2006tf}.
  The second advantage of supergroup WZNW models is the
  existence of an action principle. The latter is particularly
  powerful when it comes to the construction of local
  logarithmic field theories, a subject that has been notoriously
  hard to address for logarithmic extensions of minimal
  models. In fact, we have seen in section 5.3 that the action
  leads to a rigorous tool for constructing bulk correlation
  functions. As such, it has already been exploited in the
  construction of correlation functions for the $GL(1|1)$
  WZNW model \cite{Schomerus:2005bf}.
\medskip

  The present work admits natural extensions in several directions.
  Among these, the problem of finding concrete expressions for the
  full correlation functions or, at least, conformal blocks is
  probably the most urgent. Another issue of considerable significance
  is the extension of our ideas to world-sheets with boundaries or, in
  string theory language, the discussion of D-branes. In this context
  it seems necessary to obtain a better handle on modular transformation
  properties of characters, including those of irreducible atypical
  representations \cite{Semikhatov:2003uc}. We hope that our work will
  be helpful in deriving new character formulas along the lines of
  \cite{Gotz:2006qp,Saleur:2006tf}. It would also be interesting to
  work out in greater detail the solution of WZNW models with
  non-trivial modular invariants.
\smallskip

  In order to acquire more experience with supersymmetric $\sigma$-models
  and for various applications it would be desirable to extend our study
  to supercoset models. In contrast to bosonic models, there is considerably more
  freedom in choosing how to gauge. Besides gauging the standard adjoint
  action, as is done in \cite{Gawedzki:1988hq,Gawedzki:1989nj,Karabali:1989au,%
  Karabali:1990dk} there are many cases in which purely one-sided cosets
  are known or believed to be conformally invariant \cite{Metsaev:1998it,%
  Berkovits:1999zq,Kagan:2005wt,Babichenko:2006uc}. Those latter cases are
  relevant for the description of AdS-spaces, projective superspaces and
  even flat Minkowski space. It
  is worth noting that the harmonic analysis on coset models $G/H$ with $H$
  a bosonic subgroup acting from the right, $g\sim gh$, can easily be obtained
  from our results, see section \ref{sc:HarmAn} and especially
  eq.~\eqref{eq:SpectrumMixed} (cmp.\ also \cite{Zhang:2006}). All the
  additional input required is the branching of $\g_0$-modules into
  $\h$-modules. Even before carrying out any such decomposition explicitly,
  we may conclude from eq.~\eqref{eq:SpectrumMixed} that the resulting
  $\g$-modules are all projective. This particularly applies to all
  generalized symmetric spaces which are relevant for the description
  of AdS-spaces. Let us stress, however, that cosets $G/H$ by some non-trivial
  {\em super}group $H$ may behave differently. In fact, some simple examples
  show how even atypical irreducibles may emerge in their spectrum.
\smallskip

  Apart from these structural and conceptual issues we also expect
  our work to have concrete implications, e.g.\ in string theory. Let us
  recall that it is not difficult to write down {\em classical} $\sigma$-models which
  can be used to describe string theory on AdS-spaces with various types
  of background fluxes for instance \cite{Metsaev:1998it,Berkovits:1999im}.
  But for a long time it has not been clear how to {\em quantize} these
  field theories while keeping the target space supersymmetry manifest.
  It was only recently that the pure spinor approach closed this gap to
  some extent \cite{Berkovits:2000fe,Grassi:2001ug}. Although substantial
  progress has been made on certain aspects of the pure spinor formulation,
  there exist a variety of open conceptual issues, in particular when
  curved backgrounds are involved. It was proposed to overcome some of
  them through a reformulation in terms of supergroup WZNW models 
  \cite{Grassi:2003kq}. The ideas presented above may help to gain 
  more control over the relevant models.  
\smallskip 
 
  For a complete picture we also need to solve WZNW models beyond
  Lie supergroups of type~I. These include, in particular, supergroups 
  of type II where the fermions occur in a {\em single}
  multiplet of the bosonic subgroup. Structurally, type~II implies that
  there is no natural $\Integer$-grading anymore which is consistent with
  the intrinsic $\Integer_2$-grading of the underlying Lie superalgebra.
  This issue concerns the two series $B(m,n)=osp(2m+1|2n)$ and
  $D(m,n)=osp(2m|2n)$ of Kac' classification \cite{Kac:1977em} which
  e.g.\ constitute the isometries of superspheres $S^{2n+m-1|2n}$
  \cite{Read:2001pz,Jacobsen:2005uw}. Moreover, these series include the
  special cases $D(2,1;\alpha)$ and $D(n+1,n)$ which have been shown
  \cite{Babichenko:2006uc} to have similarly exciting properties as
  $A(n,n)=psl(n+1|n+1)$ \cite{Bershadsky:1999hk}. Let us note that
  the WZNW models based on the family of exceptional Lie superalgebras
  $D(2,1;\alpha)$ are also relevant for a manifestly supersymmetric
  description of string backgrounds involving $AdS_3\times S^3\times S^3$.
  Since type~II superalgebras do not admit a canonical (covariant) split
  of fermionic coordinates into holomorphic and antiholomorphic degrees
  of freedom, they require a rather significant extension of the above
  analysis. Incidently, the same is true for the representation theory
  of type~II superalgebras which is considerably more complicated than
  in the type~I case \cite{Kac1978:MR519631}. We hope to return to these
  issues in future work.

\subsubsection*{Acknowledgements}

  It is a great pleasure to thank Nathan Berkovits, Thomas Creutzig, Matthias
  Gaberdiel, Gerhard G\"otz, Andreas Ludwig, Eric Opdam, J{\o}rgen Rasmussen,
  Alice Rogers, Ingo Runkel, Hubert Saleur, Alexei Semikhatov, Vera Serganova, Arvind 
  Subramaniam and Anne Taormina for useful and inspiring discussions. Thomas Quella 
  acknowledges
  the warm hospitality at the KITP in Santa Barbara during an intermediate stage
  of this work. This research was supported in part by the EU Research Training
  Network grants ``Euclid'' (contract number HPRN-CT-2002-00325),
  ``ForcesUniverse'' (contract number MRTN-CT-2004-005104), ``Superstring
  Theory'' (contract number MRTN-CT-2004-512194), by the PPARC rolling grant
  PP/C507145/1, and by the National Science Foundation under Grant
  no.\ PHY99-07949. Part of this work has been performed while Thomas Quella
  was working at King's College London, funded by a PPARC postdoctoral
  fellowship under reference PPA/P/S/2002/00370.

\def\cprime{$'$} \def\cprime{$'$}
\providecommand{\href}[2]{#2}\begingroup\raggedright\endgroup


\end{document}